# Investigative monitoring of pesticide and nitrogen pollution sources in a complex multi-stressed catchment: the Lower Llobregat River basin case study (Barcelona, Spain)

Cristina Postigo[1*], Antoni Ginebreda[1*], Maria Vittoria Barbieri[1], Damià Barceló[1,2], Jordi Martin[3], Agustina de la Cal[3], Maria Rosa Boleda[3], Neus Otero[4,5], Raul Carrey[4], Vinyet Solá[6], Enric Queralt[6], Elena Isla[7], Anna Casanovas[7], Gemma Frances[7], Miren López de Alda[1]

[1] *Department of Environmental Chemistry, Institute of Environmental Assessment and Water Research (IDAEA-CSIC), Carrer de Jordi Girona 18-26, 08034, Barcelona, Spain*

[2] *Catalan Institute for Water Research (ICRA), Emili Grahit, 101, Edifici H2O, Parc Científic i Tecnològic de la Universitat de Girona, 17003 Girona, Spain.*

[3] *Aigües de Barcelona, Empresa Metropolitana de Gestió del Cicle Integral de l'Aigua, S.A. Carrer de General Batet 1-7, 08028 Barcelona, Spain.*

[4] *Grup MAiMA, SGR Mineralogia Aplicada, Geoquímica i Geomicrobiologia, Departament de Mineralogia, Petrologia i Geologia Aplicada, Facultat de Ciències de la Terra, Universitat de Barcelona (UB), Carrer de Martí i Franquès s/n, 08028 Barcelona, Spain, i Institut de Recerca de l'Aigua (IdRA), UB.*

[5] *Serra Húnter Fellowship, Generalitat de Catalunya, Spain.*

[6] *Comunitat d'Usuaris d'Aigües de la Vall Baixa i del Delta del Llobregat (CUADLL), Carrer de Pau Casals 14-16, local, 08820, El Prat de Llobregat, Spain.*

[7] *Parc Agrari del Baix Llobregat, Can Comas, Camí de la Rivera, s/n, 08820 El Prat de Llobregat, Spain;*

*Corresponding authors:

Cristina Postigo (0000-0002-7344-7044) cprgam@cid.csic.es
Antoni Ginebreda (0000-0003-4714-2850) agmgam@cid.csic.es
Institute of Environmental Assessment and Water Research (IDAEA-CSIC)
Department of Environmental Chemistry
C/ Jordi Girona 18-26, 08034 Barcelona, Spain.
Tel: +34-934-006-100, Fax: +34-932-045-904

## Abstract

The management of the anthropogenic water cycle must ensure the preservation of the quality and quantity of water resources and their careful allocation to the different uses. Protection of water resources requires the control of pollution sources that may deteriorate them. This is a challenging task in multi–stressed catchments. This work presents an approach that combines pesticide occurrence patterns and stable isotope analyses of nitrogen ($\delta^{15}N$–$NO_3^-$, $\delta^{15}N$–$NH_4^+$), oxygen ($\delta^{18}O$–$NO_3^-$), and boron ($\delta^{11}B$) to discriminate the origin of pesticides and nitrogen–pollution to tackle this challenge. The approach has been applied to a Mediterranean sub–catchment subject to a variety of natural and anthropogenic pressures. Combining the results from both analytical approaches in selected locations of the basin, the urban/industrial activity was identified as the main pressure on the quality of the surface water resources, and to a large extent also on the groundwater resources, although agriculture may play also an important role, mainly in terms of nitrate and ammonium pollution. Total pesticide concentrations in surface waters were one order of magnitude higher than in groundwaters and believed to originate mainly from soil and/or sediments desorption processes and urban and industrial use, as they were mainly associated with treated wastewaters. These findings are supported by the stable isotope results, that pointed to an organic origin of nitrate in surface waters and most groundwater samples. Ammonium pollution observed in some aquifer locations is probably generated by nitrate reduction. Overall, no significant attenuation processes could be inferred for nitrate pollution. The approach presented here exemplifies the investigative monitoring envisioned in the Water Framework Directive.

## Keywords

Water pollution, nitrate; ammonium; stable isotopes; agriculture; plant protection products

## 1. Introduction

The Mediterranean Basin, especially along its coastal area, is increasingly subjected to urban, industrial, and agricultural pressures that give rise to land-use changes, growing population, and seasonal impacts due to tourism. This setting also is characterized by natural hydrological stress (water scarcity), which according to the provisions of the Intergovernmental Panel on Climate Change (IPCC) for the Mediterranean basin (IPCC, 2014) may be worsened in the forthcoming future, resulting in less water available and more unevenly distributed, as a consequence of an expected increased occurrence of extreme hydrological events (i.e., droughts and floods). This growing imbalance between water resources and demands is a reality in the Mediterranean basin and many other coastal areas. The management of the natural and anthropogenic water cycle under such circumstances may become a complex and challenging task, in which the preservation of the resources, both in quality and quantity, and the allocation to the different uses (e.g., supply for human consumption, industrial use and agriculture irrigation, and preservation of natural areas) must be carefully balanced. Meeting this increasing water demand may require the incorporation of new resources, such as (desalinated) seawater, reclaimed water from wastewater treatment plants (WWTPs), and groundwater, which requires a "fine-tuning" of the different parts of the water cycle to avoid deterioration of water resources by anthropogenic pollutants, such as nitrogen-species and pesticides. This resource management challenge, which has been comprehensively outlined by the EU H2020 project WATERPROTECT (www.water-protect.eu), requires the collaboration of policymakers, water authorities, operators, and water users. The governance of the water cycle must fulfill both the environmental and human health requirements. The ecological status of surface waters is governed by the Water Framework Directive (WFD) (Directive 2000/60/EC) (EC, 2000), and its daughter Directive 2008/105/EC (EC, 2008), updated by Directive 2013/39/EU (EC, 2013). These regulations set specific limits (Environmental Quality Standards, EQS) for 45 priority substances, among which 24 are pesticides or biocides. New candidates for potential inclusion in the WFD list of priority substances are gathered in the so-called 'Watch List' (EC, 2018).

The WFD addresses groundwater quality in Directive 2006/118/EC (EC, 2006), setting maximum levels for nitrates (50 mg/L) and pesticides (0.1 μg/L per substance and 0.5 μg/L for total pesticides). These parametric values also hold for any water intended for human consumption (EC, 1998; EC, 2015). Nitrogen pollution of surface and groundwater from agricultural sources is also addressed by the Nitrates Directive (Directive 91/676/EEC) under the umbrella of the WFD.

The identification of pollution sources is crucial to protect water resources and ensure the good ecological status of surface water bodies as well as the good chemical status of both surface and groundwater bodies. The current analytical instrumentation allows reliable detection of organic pollutants, such as pesticides, at relevant environmental concentrations (from pg to μg/L levels). The widespread occurrence of pesticides in the environment may be linked to their application in agriculture and diffuse release from soil and sediments (Barbieri et al., 2019) and to point contamination sources such as WWTP discharges (Köck-Schulmeyer et al., 2013; Münze et al., 2017; Sutton et al., 2019), specific industries, or cleaning of pesticide application equipment. Agricultural and urban land uses are also related to the nitrogen pollution of water. The stable isotope fingerprint of $^{15}N$ in the N-species present in the water provides valuable information on their origin (anthropogenic or natural) and corresponding biochemical and physicochemical processes (nitrification, biological fixation, natural attenuation due to denitrification, or volatilization) (Nikolenko et al., 2018). For instance, inorganic fertilizers ($NH_4^+$ or $NO_3^-$) present lower $\delta^{15}N$ values (between −5‰ and +5‰) than organic sources (organic fertilizers and sewage) (between +8 ‰ and +20 ‰) (Vitòria et al., 2004). However, this initial isotopic composition can be slightly altered by physical-chemical and biochemical reactions occurring during storage and after application. The measurement of additional stable isotopes such as $\delta^{11}B$ and $\delta^{18}O$ in water can complement the aforementioned data. The distribution of $\delta^{18}O$ helps to identify nitrification processes and also discriminate $NO_3^-$ anthropogenic sources, while $\delta^{11}B$ helps to trace sewage contamination (Widory et al., 2004).

In this context, the present study aimed at integrating the information obtained from different sources to trace the origin of two major water contaminants, namely nitrogen nutrients and pesticides, in a multi-stressed river basin. For this, a typical Mediterranean area subjected to multiple pressures, *viz.*, the lower Llobregat River basin, was selected as a case study. This basin, located south of the Barcelona metropolitan area (NE Spain), is affected by urban, industrial, and agricultural activities (Sabater et al., 2012), and its aquifer system is heavily exploited, with more than 700 wells being used for drinking, agricultural, and industrial purposes (50 $Hm^3$/year), resulting in declining aquifer levels and seawater intrusion (Vázquez-Suñé et al., 2004). Pesticides and nitrogen-nutrients have been often pointed out as relevant water pollutants, exceeding in some cases the regulatory limits (Cabeza et al., 2012; Ginebreda et al., 2014; Köck-Schulmeyer et al., 2012; Masiá et al., 2015; Quintana et al., 2019; Ricart et al., 2010); however, their sources are still unknown. To trace them, a wide range of pesticides (102) were determined in a number of selected locations and various stable isotopes ($\delta^{15}N$, $\delta^{18}O$, $\delta^{11}B$) were analysed to fingerprint the N-species present in water. The information obtained with the aforementioned analytical approaches was combined with local land uses and hydrodynamics for correct data interpretation. This approach is presented as a practical implementation of the concept of investigative monitoring envisioned in the WFD.

## 2. Materials and methods

### *2.1. Study area: the lower Llobregat River basin*

The lower Llobregat River basin is a sub-catchment of the Llobregat River basin located southern to the Barcelona Metropolitan area (Figure S1 in supplementary material, SM). It covers ca. 120 $km^2$ of densely populated and highly industrialized land and includes around 30 WWTP effluent discharges along the upstream river course (Ginebreda et al., 2010). The river has a typical Mediterranean hydraulic regime, with low flows during normal conditions (around 5 $m^3/s$) and extreme events that range from absolute dryness to flooding (up to 2000 $m^3/s$) (Quintana et al., 2019) (precipitation and flow data in the sampled area are provided in Figures S2–S4). Furthermore, the area is affected by large infrastructures like highways, roads, railways, or the Barcelona harbor and airport facilities. As regards to the water cycle, the waterworks located in Sant Joan Despí supplies drinking water to approximately half of the population of the Barcelona's metropolitan area (8 $m^3/s$, serving ca. 1,5 M inhabitants) (Quintana et al., 2019) and uses water from the Llobregat River regularly mixed with groundwater when the quality or quantity of the river water is low. There are also several small drinking water treatment plants (DWTPs) in the area that supply water to selected municipalities (e.g., El Prat de Llobregat, Sant Feliu, and Sant Vicenç dels Horts) (Figure 1). Owing to the severe droughts that occurred in the past, a seawater desalination treatment plant (SWTP) (2.3 $m^3/s$ maximum) capable of supplying 22.5 % of the consumption demand of the Barcelona metropolitan area if needed was built and started operating in 2009 (Quintana et al. 2019).

Several WWTPs are located in the area, with the largest, El Prat de Llobregat (420,000 $m^3/d$), serving a daily population equivalent of 2,275,000 inhabitants. The treated effluent is partly discharged into the sea via a 3.2 km pipeline, and approximately 3% is submitted to tertiary treatment and reclaimed for several uses, including agriculture irrigation, preservation of natural protected spaces (marshlands), reduction of seawater intrusion (by ground injection), and recovery of the Llobregat river flow. The portion allocated to each of the aforementioned uses varies depending on the hydrological situation. The system, extensively described elsewhere (Custodio, 2005; Custodio, 2010; Custodio, 2017) both in terms of hydrology and quality, supports intensive agriculture activities (horticulture, vegetables, fruit trees, etc.) typically spread over multiple small extension properties located in the Agrarian Park and plays a crucial role in drinking water supply and irrigation (Figures S1 and S5). The aquifer system in the area, formed by the Lower Valley and the Delta aquifers, is under high anthropogenic pressure (urban, industrial, and agricultural) and its overexploitation has derived in seawater intrusion.

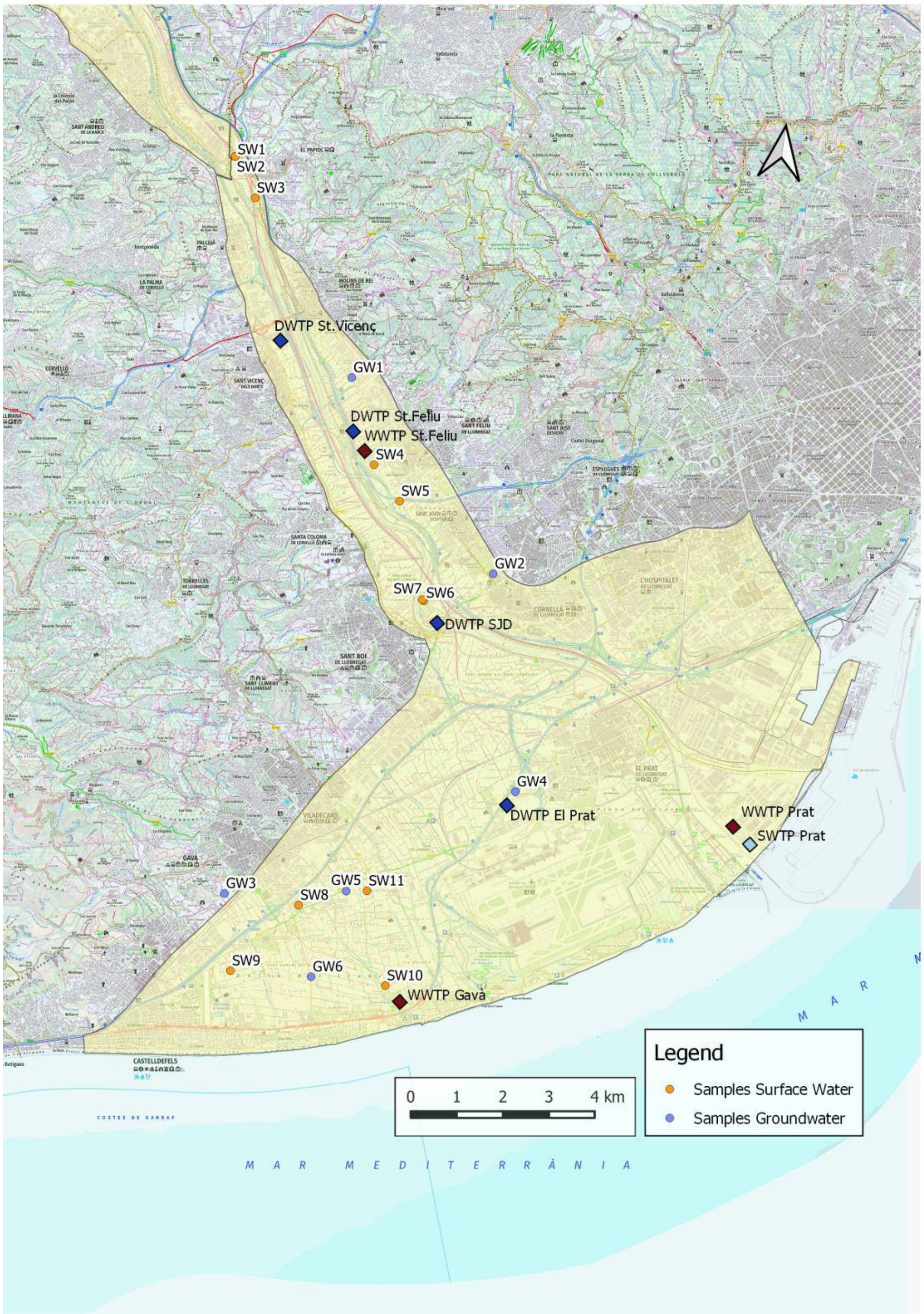

Figure 1. Location of sampling sites for pesticide and nitrogen-nutrients monitoring (DWTP: drinking water treatment plant, WWTP: wastewater treatment plant, SWTP: seawater treatment plant).

## 2.2. *Sampling details*

The sampling was designed with the contribution of all main water stakeholders in the area through a participatory monitoring approach to assess the status and main pollution sources in the area. Sampling locations, shown in Figure 1, were determined after evaluation of historical data collected in the framework of the chemical monitoring network designed by the Association of groundwater users (Comunitat d'Usuaris d'Aigües de la Vall Baixa i del Delta del Llobregat, CUADLL) and the Agrarian Park for groundwater and surface water, respectively, in the investigated area.

The sampling locations, described in detail in Table 1, are representative of the different pollution sources that may occur in the study area. Thus, agricultural activities were monitored through the sampling of irrigation channel networks that also receive agricultural field inputs (SW5, SW6, SW9, SW10, SW11), and urban and industrial activities through the sampling of highly polluted streams (SW1 and SW2) that feed an irrigation channel (SW3) (Figure S3), and wastewater treatment plant discharges (SW4 and SW8) that also feed irrigation channels. Moreover, surface water and groundwater used to produce drinking water were also monitored (SW7, GW2, GW3, and GW4). Additional groundwater samples were collected from wells used for irrigation purposes and located in the unconfined aquifers in agricultural areas (GW1 and GW5). Finally, a groundwater sample from the deep (confined) Delta aquifer (GW6) was also collected.

A total of 11 surface water locations and 6 groundwater wells were sampled in winter and summer 2019 (Table 1). For pesticide determination, water samples were collected in amber glass bottles, whereas for physical–chemical characterization and stable isotope analysis, samples were collected in plastic bottles. Grab surface water samples were collected after rinsing twice the sampling devices with the water of each sampling location. Groundwater samples were collected following the Catalan Water Agency standard operational procedure for groundwater sampling (ACA, 2015). Before collection, the monitoring wells or piezometers were pumped for 10–30 min to purge stagnant water and obtain a representative sample. Pumping time was set according to the exploitation activity in the well at the moment of sampling (in use or stopped for days). Upon collection, all samples were kept and transported under cool conditions to the analytical laboratories.

Pesticides, nitrogen–nutrients (ammonium and nitrate), and other physical–chemical parameters (e.g., turbidity, hardness, alkalinity, pH, conductivity, chloride, sulfate, iron, manganese, sodium, potassium, calcium, and magnesium) were determined in all samples collected in both sampling campaigns. The stable isotopes $^{15}N$ and $^{18}O$ of $NO_3^-$ were evaluated in all samples collected during the winter sampling campaign, whereas these and $^{15}N$ of $NH_4^+$ and $^{11}B$ were only measured in 10 selected samples of the summer campaign. Sample selection was done to further investigate the origin of nitrogen pollution in those cases that were not clear after the winter sampling campaign and to investigate ammonium origin in those locations with high ammonium levels.

Table 1. Details of the samples collected for pesticide and N–nutrients analysis (see Figure 1).

| Type | Sample ID | Full name | Descritption | UTMx | UTMy | Sampling date |
|---|---|---|---|---|---|---|
| Surface water | SW1 | Anoia | Anoia tributary (tanning industry discharges) | 416248 | 4588048 | 22/01/2019<br>27/06/2019 |
| | SW2 | Rubí Creek | Main river after Rubí creek confluence (a heavily polluted creek, partly bypass downstream, whose flow is in most part WWTPs discharges from Terrasa and Rubí) | 416249 | 4588048 | 22/01/2019<br>26/06/2019 |
| | SW3 | Infanta Channel-I | Irrigation channel (a mixture of Anoia River and Rubí Creek) | 416682 | 4587149 | 22/01/2019<br>27/06/2019 |
| | SW4 | WWTP St Feliu | Channel that collects storm runoff from highways and surrounding small creeks and | 419263 | 4581365 | 22/01/2019<br>26/06/2019 |
| | | | St Feliu WWTP effluent | | | |
| | SW5 | Infanta Channel–M | Irrigation channel (a mixture of Anoia River and Rubí Creek and potential inputs from the drainage of agricultural fields ) | 419816 | 4580574 | 22/01/2019<br>26/06/2019 |

| Type | Sample ID | Full name | Descritption | UTMx | UTMy | Sampling date |
|---|---|---|---|---|---|---|
| Surface water | SW6 | Governador tube | Irrigation channel (receives water from SW4) | 420339 | 4578416 | 22/01/2019<br>27/06/2019 |
| | SW7 | DWTP–SJD intake | Main river at the DWTP intake | 420307 | 4578441 | 22/01/2019<br>26/06/2019 |
| | SW8 | Corredora V–1 | Irrigation channel (collects Gavà–Viladecans WWTP effluent and distributes it to other irrigation channels in the network) | 417626 | 4571817 | 27/01/2019<br>25/06/2019 |
| | SW9 | Corredora V–2 | Irrigation channel (a mixture of Gavà–Viladecans WWTP effluent, groundwater from the surficial aquifer, and drainage of agricultural fields) | 416146 | 4570394 | 27/01/2019<br>25/06/2019 |
| | SW10 | Corredora V–3 | Irrigation channel (a mixture of Gavà–Viladecans WWTP effluent, groundwater from the surficial aquifer, and drainage of agricultural fields | 419506 | 4570071 | 27/01/2019<br>25/06/2019 |
| | SW11 | Corredora V–8 | Irrigation channel (a mixture of Gavà–Viladecans WWTP effluent, groundwater from the surficial aquifer, and drainage of agricultural fields | 419109 | 4572126 | 27/01/2019<br>25/06/2019 |
| Ground water* | GW1 | P–Mas Casanovas | Well used for agricultural irrigation – Low Valley aquifer (unconfined) | 418779 | 4583253 | 06/11/2018<br>26/06/2019 |
| | GW2 | SGAB–LL2 | Well used for drinking water production at the DWTP SJD | 421845 | 4579000 | 06/11/2018 |

| | | | | | |
|---|---|---|---|---|---|
| | | – Low Valley aquifer (unconfined)<br><br>Well located at the center or the main pumping area | | | 26/06/2019 |
| GW3 | SGAB–Gavà4 | Well used for drinking water production at the DWTP El Prat – Unique aquifer of the Delta aquifer system<br><br>Well located close to the aquifer margin | 416015 | 4572070 | 06/11/2018<br><br>27/06/2019 |
| GW4 | P–APSA16 | Well used for drinking water production at the DWTP El Prat – Delta deep aquifer | 422324 | 4574275 | 06/11/2018<br><br>26/06/2019 |
| GW5 | P–22CPA | Well used for agricultural irrigation – Delta surficial aquifer | 418657 | 4572120 | 27/01/2019<br><br>25/06/2019 |
| GW6 | B3A | Piezometer of the Delta surficial aquifer | 417114 | 4570981 | 27/01/2019<br><br>26/06/2019 |

*The aquifer system consists of two main aquifers:

–The Low Valley aquifer is an unconfined aquifer that is linked with the Delta deep aquifer

–The Delta aquifer, consisting of a surficial aquifer and a deep aquifer separated by a limestone layer. This layer thickens towards the coast but it does not exist close to the mountain range, and thus, in this area a unique aquifer forms (GW3).

## 2.3. *Analysis methods*

### 2.3.1. *Physical–chemical characterization of water samples*

The analysis of physical–chemical parameters was conducted at the Aigües de Barcelona laboratory. Nitrate, chloride, and sulfate were analyzed by ion chromatography (Dionex ICS–2000, Sunnyvale, CA, USA). The conductivity at 20 °C (electrometry: conductivimetric), pH (electrometry: potentiometric, glass electrode), alkalinity (acid–base potentiometric titrimetry), total hardness (complexometric titrimetry), and turbidity (nephelometry) were determined with a robòtic titrosampler (Metröhm modules 855 and 856, Herisau, Switzerland). Ammonium was analyzed using UV–VIS spectrophotometry (indophenol method) (Hewlett Packard 8453, TX, USA). All the metals tested were determined by induced coupled plasma – atomic emission spectrometry (ICP–AES) (Perkin Elmer Optima 4300 DV, Wellesley, MA, USA).

### 2.3.2. *Analysis of pesticides*

A total of 102 pesticides were analyzed in the water samples collected. The list of compounds along with the corresponding limit of quantification is provided in Table 2. Twenty–seven pesticides were determined using stir bar sorptive extraction and gas chromatography coupled to mass spectrometry (GC–MS) following previously validated methodologies (Lacorte et al., 2009; León et al., 2003). Analyses were conducted using an Agilent 7890A+ gas chromatograph coupled to a 7000C mass spectrometer (Agilent Technologies, Palo Alto, CA, USA) equipped with a TDU/CIS4 injection system (Gerstel, GmbH, Mülheuim a/d Ruhr, Germany). Polydimethylsiloxane (PDMS) coated Twister® bars (20 mm length × 0.5 mm film thickness) from Gerstel were used for the extraction of the analytes from the samples and a 5%–phenyl–methylpolysiloxane capillary column (30 m × 0.25 mm i.d. × 0.25 μm film thickness, Agilent) for their chromatographic separation.

Table 2. Target pesticides analyzed in the investigated samples and corresponding method reporting limits (MRL).

| Pesticide | CAS | Analytical method | MRL (ng/L) |
|---|---|---|---|
| 2,4,5-T | 93-76-5 | LC-MS/MS | 5 |
| 2,4,5-TP | 93-72-1 | LC-MS/MS | 5 |
| 2,4-D | 94-75-7 | LC-MS/MS | 15 |
| 2,4-DB | 94-82-6 | LC-MS/MS | 15 |
| 4,4'-DDD | 72-54-8 | GC-MS | 25 |
| 4,4'-DDE | 72-55-9 | GC-MS | 25 |
| 4,4'-DDT | 50-29-3 | GC-MS | 25 |
| Acetamiprid | 135410-20-7 | LC-MS/MS | 15 |
| Alachlor | 15972-60-8 | GC-MS | 15 |
| Aldicarb | 116-06-3 | LC-MS/MS | 25 |
| Aldrín | 309-00-2 | GC-MS | 15 |
| alpha-Endosulfan | 959-98-8 | GC-MS | 15 |
| alpha-HCH | 319-84-6 | GC-MS | 25 |
| Ametryn | 834-12-8 | GC-MS | 15 |
| Atrazine | 1912-24-9 | LC-MS/MS | 5 |
| Atrazine-desethyl (DEA) | 123948-28-7 | LC-MS/MS | 5 |
| Azoxystrobin | 131860-33-8 | LC-MS/MS | 5 |
| Bentazone | 25057-89-0 | LC-MS/MS | 5 |
| beta-Endosulfan | 33213-65-9 | GC-MS | 15 |
| beta-HCH | 319-85-7 | GC-MS | 25 |
| Bromoxynil | 1689-84-5 | LC-MS/MS | 25 |
| Carbaryl | 63-25-2 | LC-MS/MS | 15 |
| Carbendazim | 10605-21-7 | LC-MS/MS | 25 |
| Carbofuran | 1563-66-2 | LC-MS/MS | 5 |
| Chlorfenvinphos | 470-90-6 | LC-MS/MS | 5 |
| Chlorotoluron | 15545-48-9 | LC-MS/MS | 5 |
| Chloroxuron | 1982-47-4 | LC-MS/MS | 5 |
| Chlorpyrifos | 208-622-6 | GC-MS | 15 |
| Crimidine | 21725-46-2 | LC-MS/MS | 5 |
| Cyanazine | 121552-61-2 | LC-MS/MS | 5 |
| Cyprodinil | 1563-66-2 | LC-MS/MS | 5 |
| DIA (deisopropyl-atrazine) | 1007-28-9 | LC-MS/MS | 25 |
| Diazinon | 333-41-5 | LC-MS/MS | 5 |
| Dichlobenil | 1194-65-6 | GC-MS | 15 |
| Dieldrin | 60-57-1 | GC-MS | 15 |
| Dimethoate | 60-51-5 | LC-MS/MS | 15 |
| Diuron | 330-54-1 | LC-MS/MS | 15 |
| EPTC | 759-94-4 | LC-MS/MS | 5 |
| Ethofumesate | 26225-79-6 | GC-MS | 15 |
| Fenitrothion | 122-14-5 | GC-MS | 15 |
| Fenuron | 101-42-8 | LC-MS/MS | 25 |

Table 2. (continued)

| Pesticide | CAS | Analytical method | MRL (ng/L) |
|---|---|---|---|
| Flufenacet | 142459-58-3 | LC-MS/MS | 5 |
| Fluroxypyr | 69377-81-7 | LC-MS/MS | 25 |
| Heptachlor | 76-44-8 | GC-MS | 15 |
| Heptachlor-epoxide | 1024-57-3 | GC-MS | 15 |
| Imidacloprid | 138261-41-3 | LC-MS/MS | 5 |
| Ioxynil | 1689-83-4 | LC-MS/MS | 25 |
| Irgarol | 28159-98-0 | LC-MS/MS | 5 |
| Isoprocarb | 2631-40-5 | LC-MS/MS | 5 |
| Isoproturon | 34123-59-6 | LC-MS/MS | 15 |
| Lindane | 58-89-9 | GC-MS | 15 |
| Linuron | 330-55-2 | LC-MS/MS | 15 |
| MCPA | 94-81-5 | LC-MS/MS | 25 |
| MCPB | 94-81-5 | LC-MS/MS | 5 |
| MCPP (Mecoprop) | 93-65-2 | LC-MS/MS | 15 |
| Metalaxyl | 57837-19-1 | LC-MS/MS | 5 |
| Metamitron | 41394-05-2 | LC-MS/MS | 25 |
| Metazachlor | 67129-08-2 | LC-MS/MS | 15 |
| Methabenzthiazuron | 18691-97-9 | LC-MS/MS | 5 |
| Methiocarb | 2032-65-7 | LC-MS/MS | 5 |
| Methomyl | 16752-77-5 | LC-MS/MS | 25 |
| methyl-Parathion | 298-00-0 | GC-MS | 25 |
| Metobromuron | 3060-89-7 | LC-MS/MS | 5 |
| Metolachlor | 51218-45-2 | LC-MS/MS | 15 |
| Metolaclor-ESA | 171118-09-5 | LC-MS/MS | 25 |
| Metoxuron | 19937-59-8 | LC-MS/MS | 5 |
| Metribuzin | 21087-64-9 | LC-MS/MS | 5 |
| Mevinphos-(E+Z) | 7786-34-7 | LC-MS/MS | 5 |
| Molinate | 2212-67-1 | GC-MS | 25 |
| Monolinuron | 1746-81-2 | LC-MS/MS | 5 |
| Paraoxon-ethyl | 311-45-5 | LC-MS/MS | 5 |
| Parathion | 56-38-2 | GC-MS | 25 |
| Pencycuron | 66063-05-6 | LC-MS/MS | 5 |
| Pendimetalin | 40487-42-1 | GC-MS | 25 |
| Pentachlorophenol | 87-86-5 | LC-MS/MS | 25 |
| Pethoxamid | 106700-29-2 | LC-MS/MS | 5 |
| Pirimicarb | 23103-98-2 | GC-MS | 25 |
| Prochloraz | 67747-09-5 | LC-MS/MS | 25 |
| Prometon | 1610-18-0 | LC-MS/MS | 5 |
| Prometryn | 7287-19-6 | GC-MS | 15 |
| Propanil | 709-98-8 | GC-MS | 15 |
| Propazine | 139-40-2 | LC-MS/MS | 5 |
| Propham | 122-42-9 | LC-MS/MS | 15 |

Table 2. (continued)

| Pesticide | CAS | Analytical method | MRL (ng/L) |
|---|---|---|---|
| Propiconazole | 60207-90-1 | LC-MS/MS | 15 |
| Propoxur | 114-26-1 | LC-MS/MS | 5 |
| Propyzamide | 23950-58-5 | LC-MS/MS | 5 |
| Prosulfocarb | 52888-80-9 | LC-MS/MS | 25 |
| Sebuthylazine | 7286-69-3 | LC-MS/MS | 5 |
| Simazine | 122-34-9 | LC-MS/MS | 5 |
| Sulcotrione | 99105-77-8 | LC-MS/MS | 5 |
| Tebuconazole | 107534-96-3 | LC-MS/MS | 5 |
| Tebuthiuron | 34014-18-1 | LC-MS/MS | 5 |
| Terbuthylazina-2-hydroxy | 66753-07-9 | LC-MS/MS | 25 |
| Terbuthylazine | 5915-41-3 | LC-MS/MS | 5 |
| Terbutilazina-desethyl | 30125-63-4 | LC-MS/MS | 5 |
| Terbutryn | 886-50-0 | LC-MS/MS | 5 |
| Thiabendazole | 148-79-8 | LC-MS/MS | 25 |
| Thiachloprid | 111988-49-9 | LC-MS/MS | 5 |
| Thiamethoxam | 153719-23-4 | LC-MS/MS | 25 |
| Thiobencarb | 28249-77-6 | GC-MS | 15 |
| Triclopyr | 55335-06-3 | LC-MS/MS | 5 |
| Trifluralin | 1582-09-8 | GC-MS | 15 |

The remaining pesticides were analyzed using a fully automated method based on on–line solid–phase extraction and liquid chromatography–tandem mass spectrometry determination (SPE–LC–MS/MS). Analyses were conducted using an Advance™ UHPLC$^{OLE}$ system coupled to EVOQ Elite mass spectrometer (Bruker Daltonics Inc, Fremon, CA). Sample preconcentration was done on a YMC C18 trap column (30 mm × 2.1 mm i.d., particle size 10 µm), while chromatographic separation was done on a YMC C18 column (100 mm × 2.1 mm i.d., particle size 2 µm) (both from Bruker). Further details on the analytical method used and its performance are published in Quintana et al. (2019).

### 2.3.3. *Stable isotope analysis*

Samples for stable isotope analysis were filtered with a 0.2 μm polytetrafluoroethylene (PTFE) filter (Millipore®, Merck), and preserved at +4°C until their analysis. The $\delta^{15}N-NO_3^-$ and the $\delta^{18}O-NO_3^-$ analyses were performed following the Cd reduction method (McIlvin and Altabet, 2005) with an automatic pre-concentrator (Pre-Con, Thermo Scientific) coupled to an isotope-ratio mass spectrometer (IRMS) (Finnigan MAT-253, Thermo Scientific). The analysis of $\delta^{15}N-NH_4^+$ was performed following the hypobromite method described by Zhang et al. (Zhang et al., 2007). The analysis of $\delta^{11}B$ was performed using a high-resolution inductively coupled plasma mass spectrometer (HR-ICP-MS) Element XR (Thermo Scientific) following a previously published method (Gäbler and Bahr, 1999).

According to Coplen (Coplen, 2011), several international and laboratory standards were interspersed among samples for normalization of analyses. Three international standards (USGS 32, 34 and 35) and one internal laboratory standard (CCIT-IWS ($\delta^{15}N$ = +16.9 ‰ and $\delta^{18}O$ = +28.5 ‰)) were employed to correct $\delta^{15}N-NO_3^-$ and $\delta^{18}O-NO_3^-$ values. Regarding $\delta^{15}N-NH_4^+$, two international standards (USGS-25 and IAEA-N2) and two internal laboratory standards (CCIT-IWS-$NO_2^-$ ($\delta^{15}N$ = -28.5 ‰), and CCIT-IWS-$NH_4^+$ ($\delta^{15}N$= -0.8 ‰)) were employed. For $\delta^{11}B$, values were corrected using the international standard NBS-951. Isotopic results were expressed as delta ($\delta$) per mil relative to established international standards: N-atmospheric international standard for $\delta^{15}N$, Vienna Standard Mean Ocean Water (VSMOW) for the $\delta^{18}O$, and NIST-951 in the case of $\delta^{11}B$. Samples for isotopic analyses of $\delta^{15}N-NO_3^-$, $\delta^{18}O-NO_3^-$ and $\delta^{15}N-NH_4^+$ were prepared at the laboratory of the MAiMA-UB research group and analyzed at the Centres Científics i Tecnològics of the Universitat de Barcelona (CCiT-UB), whereas $\delta^{11}B$ analysis was conducted by labGEOTOP (Laboratory of Elemental and Isotopic Geochemistry for Petrological Applications) of the Institute of Earth Sciences Jaume Almera of the Spanish Scientific Research Council (ICTJA-CSIC).

## 3. Results and discussion

### *3.1. Physical–chemical characterization: salinity and nitrogen–compounds concentration*

The results of the physical–chemical characterization of the investigated waters are provided in Tables 3 and 4 and depicted for selected parameters (chloride, conductivity, nitrate, and ammonium) in Figure 2. All surface waters were saline (conductivity values within the range 1455–2978 μS/cm), even at the point of abstraction for drinking water production. Chloride ($Cl^-$) concentrations ranged between 334 and 444 mg/L in summer and from 244 to 668 mg/L in winter. There was not a fixed seasonal pattern of chloride concentrations, which varied among the investigated locations. While $Cl^-$ in SW was higher in summer (334–421 mg/L) than in winter (244–290 mg/L) upstream the DWTP SJD intake, that supplies drinking water to Barcelona, the opposite pattern was observed in the water samples collected at the irrigation and drainage channels located in the agricultural areas nearby the coast (SW8 to SW11) (a mixture of groundwater and treated wastewater), where $Cl^-$ concentrations reached 609–668 mg/L in winter. Overall, $Cl^-$ in groundwater also increased with proximity of the well to the seaside, showing a certain degree of saline intrusion, in particular in GW6, the well of the Delta aquifer closest to the coast. $Cl^-$ levels measured in surface water and groundwater, that surpassed in most cases the parametric value of 250 mg/L set in the EU Drinking Water Directive, definitely harm crop production in the area. As expected, $Cl^-$ concentrations were in line with the conductivity values (1180–2978 μS/cm in all locations except in GW6, where conductivity rose to 15000 μS/cm in winter and 27470 μS/cm in summer due to seawater intrusion).

Table 3. Physical–chemical characterization of the surface and groundwater samples analyzed in winter 2019.

| Type | Sample ID | Descriptor | Nitrate [mg $NO_3^-$/L] | Ammonium [mg $NH_4^+$/L] | Chloride [mg $Cl^-$/L] | Sulfate [mg $SO_4^{-2}$/L] | Conductivity [µS/cm] | pH | Alkalinity [mg $CaCO_3$/L] | Hardness [mg $CaCO_3$/L] | Turbidity [FNU] |
|---|---|---|---|---|---|---|---|---|---|---|---|
| Surface water | SW1 | ANOIA | 0.5 | 34 | 281 | 157 | 1725 | 7.9 | 403 | 506 | 13 |
| | SW2 | RUBI | 16 | 23 | 247 | 156 | 1504 | 8.0 | 308 | 461 | 28 |
| | SW3 | INF-I | 16 | 23 | 246 | 154 | 1501 | 7.9 | 307 | 371 | 29 |
| | SW4 | WWTP | 8.4 | 12 | 290 | 224 | 1789 | 7.6 | 367 | 456 | 3.4 |
| | SW5 | INF-M | 28 | 16 | 244 | 147 | 1455 | 7.2 | 242 | 439 | 45 |
| | SW6 | GOV | 13 | 12 | 278 | 205 | 1691 | 7.8 | 340 | 418 | 45 |
| | SW7 | DWTP | 20 | 0.82 | 275 | 214 | 1583 | 7.9 | 242 | 557 | 7.9 |
| | SW8 | V-1 | 27 | 1.7 | 668 | 235 | 2776 | 8.1 | 328 | 579 | 1.2 |
| | SW9 | V-2 | 30 | 0.43 | 609 | 214 | 2815 | 8.0 | 323 | 574 | 1.3 |
| | SW10 | V-3 | 22 | 0.53 | 661 | 259 | 2978 | 8.2 | 363 | 659 | 16 |
| | SW11 | V-8 | 35 | 2.9 | 662 | 238 | 2919 | 8.1 | 328 | 630 | 3.6 |
| Ground | GW1 | MCAS | 11 | <0.15 | 177 | 154 | 1180 | 7.5 | 251 | 363 | 1.1 |
| water | GW2 | LL2 | 41 | <0.15 | 294 | 212 | 1727 | 7.4 | 339 | 749 | 0.2 |
| | GW3 | GAVA4 | 95 | <0.15 | 198 | 140 | 1429 | 7.4 | 329 | 585 | 0.35 |
| | GW4 | APSA16 | 13 | <0.15 | 683 | 350 | 2912 | 7.2 | 323 | 840 | 0.47 |
| | GW5 | 22CPA | 0.7 | 57 | 166 | 556 | 2323 | 7.8 | 580 | 860 | 130 |
| | GW6 | B3A | 2.2 | 34 | 5000 | 21.8 | 15000 | 8.0 | 874 | 1770 | 16 |
| *EU parametric values for waters intended for human consumption (EC, 1998)* | | | 50 | 0.5 | 250 | 250 | 2500 | 6.5-9.5 | na | na | 1 |

Table 4. Physical–chemical characterization of the surface and groundwater samples analyzed in summer 2019.

| Type | Sample ID | Descriptor | Nitrate [mg $NO_3^-$/L] | Ammonium [mg $NH_4^+$/L] | Chloride [mg $Cl^-$/L] | Sulfate [mg $SO_4^{-2}$/L] | Conductivity [µS/cm] | pH | Alkalinity [mg $CaCO_3$/L] | Hardness [mg $CaCO_3$/L] | Turbidity [FNU] |
|---|---|---|---|---|---|---|---|---|---|---|---|
| Surface water | SW1 | ANOIA | 5.9 | 0.23 | 347 | 470 | 2196 | 8.2 | 332 | 760 | 20 |
| | SW2 | RUBI | 26 | 10 | 358 | 159 | 1796 | 8.1 | 247 | 439 | 5.2 |
| | SW3 | INF-I | 13 | 8.6 | 364 | 282 | 2009 | 8.0 | 293 | 508 | 23 |
| | SW4 | WWTP | 19 | 4.2 | 421 | 217 | 2085 | 7.6 | 291 | 512 | 1.5 |
| | SW5 | INF-M | 23 | 12 | 334 | 162 | 1733 | 8.0 | 250 | 463 | 5.4 |
| | SW6 | GOV | 21 | 7.7 | 386 | 238 | 2001 | 7.9 | 281 | 542 | 4.9 |
| | SW7 | DWTP | 3.9 | <0.15 | 344 | 153 | 1627 | 8.3 | 216 | 395 | 13 |
| | SW8 | V-1 | 35 | 1.5 | 425 | 144 | 1987 | 7.9 | 257 | 391 | 5.5 |
| | SW9 | V-2 | 44 | 0.29 | 426 | 144 | 1983 | 7.9 | 249 | 424 | 1.5 |
| | SW10 | V-3 | 7.7 | 0.07 | 444 | 158 | 2000 | 8.1 | 263 | 389 | 59 |
| | SW11 | V-8 | 43 | 0.95 | 434 | 144 | 2010 | 8.1 | 255 | 467 | 13 |
| Ground | GW1 | MCAS | 14 | <0.15 | 212 | 160 | 1316 | 7.7 | 241 | 425 | 0.31 |
| water | GW2 | LL2 | 19 | <0.15 | 263 | 186 | 1633 | 7.7 | 339 | 583 | 1.0 |
| | GW3 | GAVA4 | 100 | <0.15 | 194 | 142 | 1467 | 7.6 | 340 | 588 | 0.27 |
| | GW4 | APSA16 | 17 | <0.15 | 433 | 396 | 2468 | 7.4 | 395 | 845 | 1.2 |
| | GW5 | 22CPA | <1.5 | 65 | 171 | 688 | 2389 | 7.7 | 529 | 973 | 130 |
| | GW6 | B3A | 6.0 | 37 | 10770 | 19 | 27470 | 7.9 | 862 | 2000 | 15 |
| *EU parametric value in waters intended for human consumption (EC, 1998)* | | | 50 | 0.5 | 250 | 250 | 2500 | 6.5-9.5 | na | na | 1 |

na – no parametric value set in the legislation

Table 4. (continued).

| Type | Sample ID | Descriptor | Iron [µg/L] | Manganese [µg/L] | Sodium [mg/L] | Potassium [mg/L] | Calcium [mgL] | Magnesium [mg/L] |
|---|---|---|---|---|---|---|---|---|
| Surface water | SW1 | ANOIA | 60 | 41 | 249 | 20 | 197 | 65 |
| | SW2 | RUBI | 61 | 33 | 203 | 36 | 123 | 32 |
| | SW3 | INF-I | 83 | 70 | 238 | 24 | 134 | 42 |
| | SW4 | WWTP | 31 | 56 | 236 | 38 | 147 | 35 |
| | SW5 | INF-M | 68 | 64 | 189 | 35 | 129 | 34 |
| | SW6 | GOV | 64 | 42 | 236 | 36 | 151 | 40 |
| | SW7 | DWTP | 15 | 21 | 168 | 35 | 107 | 31 |
| | SW8 | V-1 | 62 | 77 | 241 | 39 | 107 | 30 |
| | SW9 | V-2 | 51 | 27 | 239 | 38 | 117 | 32 |
| | SW10 | V-3 | 72 | 97 | 259 | 34 | 103 | 32 |
| | SW11 | V-8 | 75 | 60 | 283 | 42 | 129 | 35 |
| Groundwater | GW1 | MCAS | 14 | <2 | 120 | 22 | 119 | 31 |
| | GW2 | LL2 | 7 | <2 | 158 | 12 | 154 | 48 |
| | GW3 | GAVA4 | 5 | <2 | 112 | 6 | 156 | 45 |
| | GW4 | APSA16 | 43 | 14 | 217 | 11 | 221 | 71 |
| | GW5 | 22CPA | 156 | 280 | 151 | 56 | 282 | 65 |
| | GW6 | B3A | 1375 | 50 | 5675 | 187 | 59 | 450 |
| *EU parametric value in waters intended for human consumption (EC, 1998)* | | | 200 | 50 | 200 | na | na | na |

na – no parametric value set in the legislation

Nitrate ($NO_3^-$) concentrations were, with one exception (GW3), all below the parametric value of 50 mg/L set for drinking water and groundwater quality preservation. Overall, nitrate concentrations were slightly higher in summer than in winter, with smaller differences observed in groundwater. The major exception was observed in GW2, whose nitrate concentration in winter doubled the concentration measured in summer. Ammonium ($NH_4^+$) concentrations ranged from <0.15 to 65 mg/L. All surface waters except SW1 (a Llobregat tributary diverted into an irrigation channel) and SW7 (the point of abstraction for drinking water production) in summer exceeded the drinking water threshold (0.5 mg $NH_4^+$/L) in both investigated periods. Contrary to $NO_3^-$, ammonium concentrations in summer were lower than in winter. As for groundwater, the wells used for drinking water production (GW2, GW3, and GW4) did not present detectable ammonium levels. In contrast and except for GW1 (<0.015 mg $NH_4^+$/L), high ammonium concentrations (24–65 mg $NH_4^+$/L) were detected in the remaining investigated wells (GW5 and GW6). The origin of ammonium and nitrate concentrations observed are discussed in section 3.3 of the present manuscript.

It is important to highlight that if pollution sources are constant throughout the year, concentrations in winter could be lightly diluted by a slightly higher flow in the sampling locations (Figures S3 and S4). In the main river, the daily average river flow was 1.5 higher in winter than in summer, and in summer, the average flow diverted into the irrigation channel (SW3) was 1.2 higher in winter than in summer. Furthermore, the precipitation event occurring two days before sampling SW1–SW7 may have increased flow in SW4 due to storm runoff and transport pollutants from other areas into the sampling locations (Figure S2).

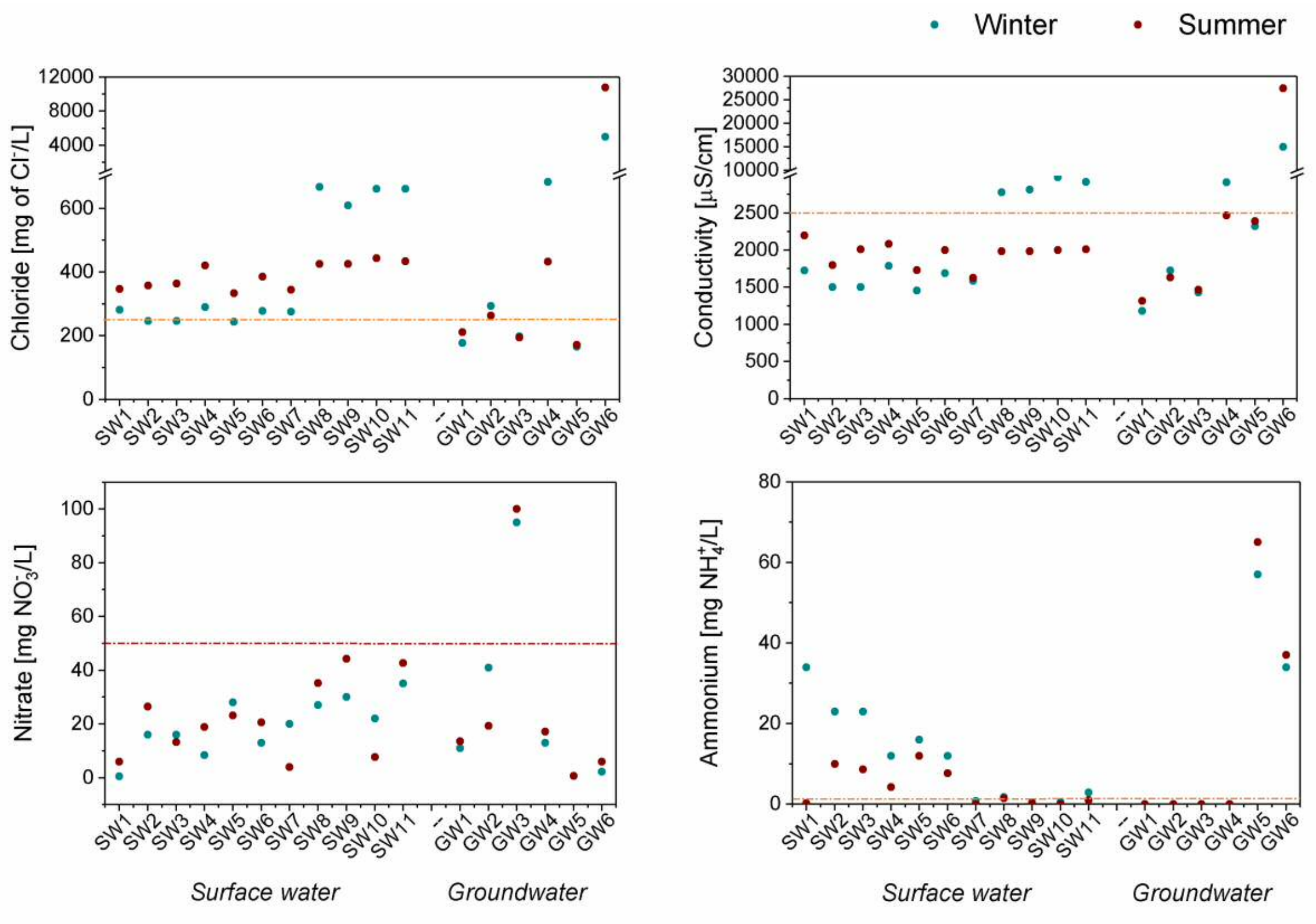

Figure 2. Chloride, conductivity, nitrate, and ammonium concentrations in the investigated waters. The dotted line indicates the parametric value set for each parameter in the EU Drinking Water Directive.

### *3.2. Occurrence and source of pesticide pollution*

From the 102 targeted pesticide compounds, only 28 were detected in the investigated waters (22 in each sampling campaign) (Figure 3 and Tables 5 and 6). Pesticide pollution was more ubiquitous and abundant in surface water than in groundwater. Maximum total pesticide concentrations in surface waters reached 1.3 µg/L and 1.9 µg/L in winter and summer, respectively. The highest pesticide cumulative levels were found at SW2 (Rubí Creek that feeds together with SW1 (Anoia River) a major irrigation channel network at SW3). Pesticides levels were also relevant at SW4 (a channel that receives storm runoff and a WWTP effluent) and also at SW3 and SW5 (different locations of the irrigation channel network that serves a mixture of SW1 and SW2 for irrigation) and at SW6 (location of an irrigation channel that ends in the Llobregat River downstream the DWTP SJD and fed with SW4 water). Most of the investigated surface waters (82%) exceeded the limit of 0.5 μg/L set for total pesticides in waters intended for human consumption by the Drinking Water Directive. Pesticide concentrations in surface waters were higher in summer than in winter (1.4–2.2–x fold) in SW2–SW4, S6, and S11, while the opposite pattern was observed in SW1, and SW7–SW10 (concentrations in winter were 1.2–1.8 times higher than in summer, and up to 14.6 higher in SW7). As previously mentioned (section 3.1), a slight dilution of concentrations could be expected in winter due to differences in the water flow in the different locations (Figures S2–S4), if pesticide input is constant. However, this dilution may be compensated by higher desorption of pesticides from soils and sediments due to more frequent storm events in winter that may transport pesticides from upstream locations. Given that there is not a fixed pattern, the differences observed in currently used pesticides could be attributed to a distinct use of these compounds in the sampling periods, according to the different growing seasons of main crops in the area (artichokes in winter and tomatoes and other orchards in summer).

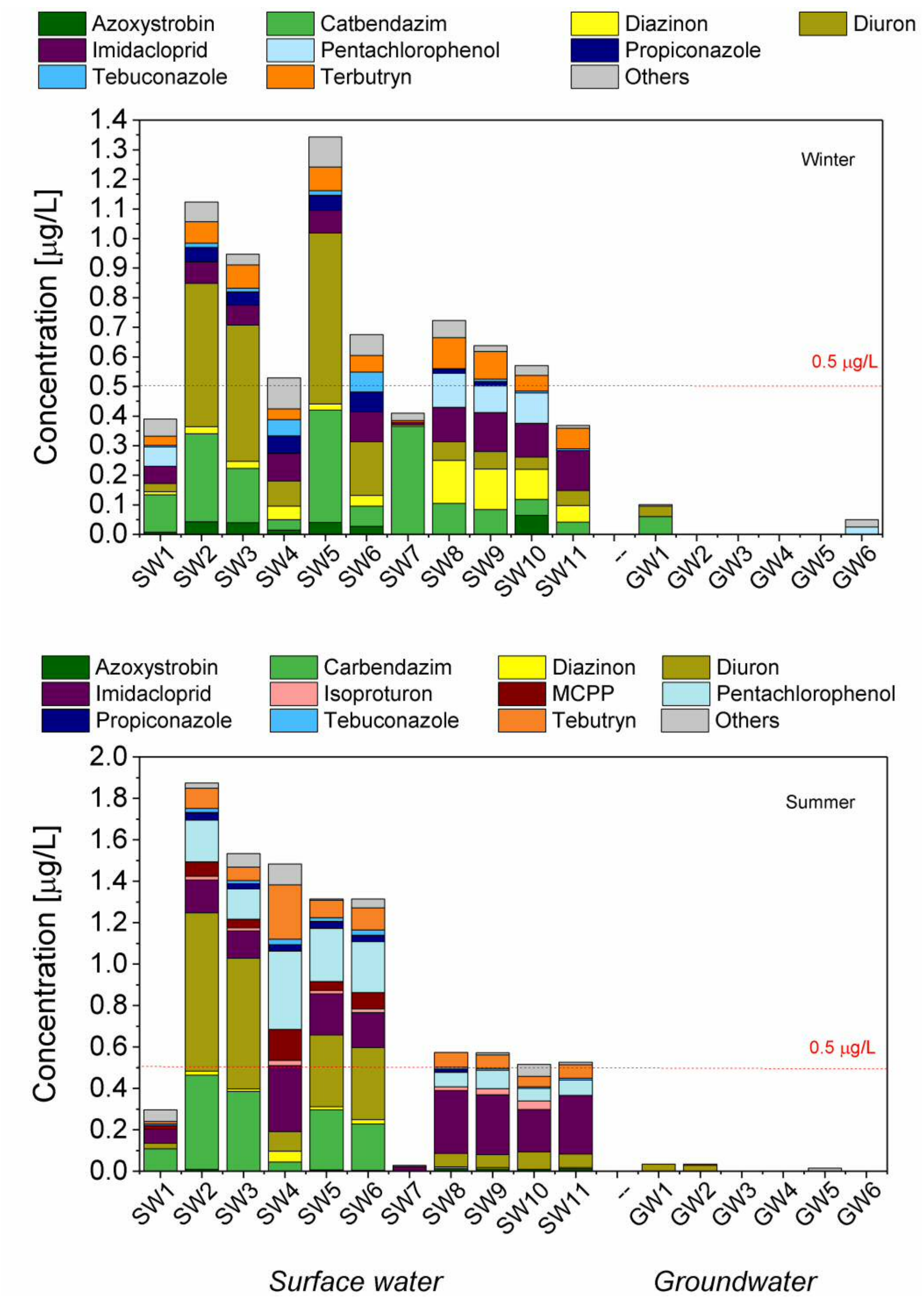

Figure 3. Concentration levels (µg/L) of the most frequently detected pesticides (>40% of the samples) or most abundant (>0.1 µg/L in at least one sample) in winter and summer. The dotted line indicates the maximum limit for total pesticide concentrations in groundwater and waters intended for human consumption. *OTHERS* in winter includes 2,4-D, bentazone, chlortoluron, cyprodinil, deisopropyl atrazine, fenuron, isoproturon, MCPA, MCPP, metalaxyl, pencycuron, and simazine. *OTHERS* in summer includes 2,4-D, bentazone, fluroxypir, lindane, MCPA, metalaxyl, propoxur, simazine, terbuthylazine-desethyl, and triclopyr.

Table 5. Pesticide concentrations (in µg/L) in the investigated water samples during the first sampling campaign (winter 2019). Concentrations above the corresponding EQS are highlighted in red.

| Sample type | Sample ID | 2,4-D | Azoxys–trobin | Ben–tazone | Carben–dazim | Chlor–toluron | Cypro–dinil | DIA* | Diazinon | Diuron | Fenuron | Imida–cloprid |
|---|---|---|---|---|---|---|---|---|---|---|---|---|
| Surface water | SW1 | <0.015 | 0.008 | <0.005 | 0.126 | <0.005 | <0.005 | 0.058 | 0.010 | 0.028 | <0.025 | 0.058 |
| | SW2 | 0.035 | 0.043 | <0.005 | 0.297 | <0.005 | <0.005 | <0.025 | 0.024 | 0.485 | <0.025 | 0.072 |
| | SW3 | 0.036 | 0.040 | <0.005 | 0.183 | <0.005 | <0.005 | <0.025 | 0.024 | 0.461 | <0.025 | 0.067 |
| | SW4 | <0.015 | 0.015 | <0.005 | 0.035 | <0.005 | 0.005 | <0.025 | 0.046 | 0.085 | 0.03 | 0.093 |
| | SW5 | 0.038 | 0.041 | <0.005 | 0.379 | <0.005 | <0.005 | <0.025 | 0.021 | 0.578 | <0.025 | 0.076 |
| | SW6 | <0.015 | 0.027 | <0.005 | 0.069 | <0.005 | <0.005 | <0.025 | 0.036 | 0.181 | 0.032 | 0.101 |
| | SW7 | 0.019 | <0.005 | <0.005 | 0.365 | 0.006 | <0.005 | <0.025 | 0.005 | <0.015 | <0.025 | 0.009 |
| | SW8 | <0.015 | <0.005 | 0.016 | 0.105 | <0.005 | <0.005 | <0.025 | 0.145 | 0.063 | <0.025 | 0.117 |
| | SW9 | <0.015 | <0.005 | 0.011 | 0.084 | <0.005 | <0.005 | <0.025 | 0.137 | 0.059 | <0.025 | 0.132 |
| | SW10 | <0.015 | 0.065 | 0.011 | 0.053 | <0.005 | <0.005 | <0.025 | 0.102 | 0.042 | <0.025 | 0.114 |
| | SW11 | <0.015 | <0.005 | 0.009 | 0.042 | <0.005 | <0.005 | <0.025 | 0.056 | 0.051 | <0.025 | 0.134 |
| Ground water | GW1 | <0.015 | <0.005 | <0.005 | 0.06 | <0.005 | <0.005 | <0.025 | <0.005 | 0.036 | <0.025 | <0.005 |
| | GW2 | <0.015 | <0.005 | <0.005 | <0.025 | <0.005 | <0.005 | <0.025 | <0.005 | <0.015 | <0.025 | <0.005 |
| | **GW3** | <0.015 | <0.005 | <0.005 | <0.025 | <0.005 | <0.005 | <0.025 | <0.005 | <0.015 | <0.025 | <0.005 |
| | **GW4** | <0.015 | <0.005 | <0.005 | <0.025 | <0.005 | <0.005 | <0.025 | <0.005 | <0.015 | <0.025 | <0.005 |
| | **GW5** | <0.015 | <0.005 | <0.005 | <0.025 | <0.005 | <0.005 | <0.025 | <0.005 | <0.015 | <0.025 | <0.005 |
| | **GW6** | <0.015 | <0.005 | <0.005 | <0.025 | <0.005 | <0.005 | <0.025 | <0.005 | <0.015 | <0.025 | <0.005 |
| **FD in SW (%)* | | 36 | 64 | 36 | 100 | 9 | 9 | 9 | 100 | 82 | 18 | |
| **FD in all (%)* | | 24 | 41 | 24 | 71 | 6 | 6 | 6 | 65 | 59 | 12 | |
| ***EQS (µg/L*)* | | – | – | – | – | – | – | – | – | 0.2 | – | |

*FD: frequency of detection in surface waters and all investigated waters

**EQS for annual average concentration or method detection limit in the case of Watch List pesticides (e.g., imidacloprid)

Table 5. (continued)

| Sample type | Sample ID | Iso–proturon | MCPA | MCPP | Meta–laxyl | Pentachlorophenol | Pency–curon | Propi–conazole | Simazine | Tebu–conazole | Ter–butryn |
|---|---|---|---|---|---|---|---|---|---|---|---|
| Surface water | **SW1** | <0.015 | <0.025 | <0.015 | <0.005 | 0.066 | <0.005 | <0.015 | <0.005 | 0.005 | 0.031 |
| | **SW2** | 0.015 | <0.025 | 0.016 | <0.005 | <0.025 | <0.005 | 0.049 | <0.005 | 0.014 | 0.073 |
| | **SW3** | <0.015 | <0.025 | <0.015 | <0.005 | <0.025 | <0.005 | 0.045 | <0.005 | 0.012 | 0.079 |
| | **SW4** | <0.015 | 0.026 | 0.038 | 0.006 | <0.025 | <0.005 | 0.059 | <0.005 | 0.055 | 0.036 |
| | **SW5** | 0.016 | 0.028 | 0.019 | <0.005 | <0.025 | <0.005 | 0.051 | <0.005 | 0.016 | 0.080 |
| | **SW6** | <0.015 | <0.025 | 0.033 | 0.005 | <0.025 | <0.005 | 0.067 | <0.005 | 0.068 | 0.056 |
| | **SW7** | <0.015 | <0.025 | <0.015 | <0.005 | <0.025 | <0.005 | <0.015 | <0.005 | <0.005 | 0.006 |
| | **SW8** | <0.015 | <0.025 | <0.015 | <0.005 | 0.115 | 0.042 | 0.015 | <0.005 | <0.005 | 0.105 |
| | **SW9** | <0.015 | <0.025 | <0.015 | <0.005 | 0.09 | 0.009 | 0.015 | <0.005 | 0.008 | 0.093 |
| | **SW10** | <0.015 | <0.025 | 0.018 | <0.005 | 0.102 | 0.005 | <0.015 | <0.005 | 0.007 | 0.052 |
| | **SW11** | <0.015 | <0.025 | <0.015 | <0.005 | <0.025 | <0.005 | <0.015 | <0.005 | 0.007 | 0.069 |
| Ground water | **GW1** | <0.015 | <0.025 | <0.015 | <0.005 | <0.025 | <0.005 | <0.015 | <0.005 | <0.005 | <0.005 |
| | **GW2** | <0.015 | <0.025 | <0.015 | <0.005 | <0.025 | <0.005 | <0.015 | 0.005 | <0.005 | <0.005 |
| | GW3 | <0.015 | <0.025 | <0.015 | <0.005 | <0.025 | <0.005 | <0.015 | <0.005 | <0.005 | <0.005 |
| | GW4 | <0.015 | <0.025 | <0.015 | <0.005 | <0.025 | <0.005 | <0.015 | <0.005 | <0.005 | <0.005 |
| | GW5 | <0.015 | <0.025 | <0.015 | <0.005 | <0.025 | <0.005 | <0.015 | <0.005 | <0.005 | <0.005 |
| | GW6 | <0.015 | <0.025 | <0.015 | <0.005 | 0.025 | <0.005 | <0.015 | <0.005 | <0.005 | <0.005 |
| **FD in SW (%)* | | 18 | 18 | 45 | 18 | 18 | 27 | 64 | 0 | 82 | 100 |
| **FD in all (%)* | | 12 | 12 | 29 | 12 | 29 | 18 | 41 | 6 | 53 | 65 |
| ***EQS (μg/L*)* | | 0.3 | – | – | – | 0.4 | – | – | 1 | – | 0.065 |

*FD: frequency of detection in surface waters and all investigated waters

**EQS for annual average concentration or method detection limit in the case of Watch List pesticides (e.g., imidacloprid)

Table 6. Pesticide concentrations (in μg/L) in the investigated water samples during the second sampling campaign (summer 2019). Concentrations above the corresponding EQS are highlighted in red.

| Sample type | Sample ID | 2.4-D | Aceta-miprid | Azoxy-strobin | Ben-tazone | Carben-dazim | Diazinon | Diuron | Fluro-xypyr | Imida-cloprid | Iso-proturon | Lindane |
|---|---|---|---|---|---|---|---|---|---|---|---|---|
| Surface water | **SW1** | <0.015 | <0.015 | <0.005 | 0.008 | 0.109 | <0.015 | 0.026 | <0.025 | 0.068 | <0.015 | 0.038 |
| | **SW2** | 0.024 | 0.028 | 0.010 | <0.005 | 0.454 | 0.020 | 0.764 | <0.025 | 0.159 | 0.017 | <0.015 |
| | **SW3** | 0.017 | <0.015 | <0.005 | <0.005 | 0.385 | 0.013 | 0.631 | 0.034 | 0.131 | 0.015 | 0.015 |
| | **SW4** | 0.033 | <0.015 | <0.005 | <0.005 | 0.045 | 0.051 | 0.096 | <0.025 | 0.318 | 0.025 | <0.015 |
| | **SW5** | <0.015 | 0.020 | 0.007 | <0.005 | 0.290 | 0.014 | 0.347 | <0.025 | 0.198 | 0.017 | <0.015 |
| | **SW6** | 0.028 | 0.021 | 0.005 | <0.005 | 0.223 | 0.022 | 0.347 | <0.025 | 0.170 | 0.017 | <0.015 |
| | **SW7** | <0.015 | <0.015 | <0.005 | <0.005 | <0.025 | <0.005 | <0.015 | <0.025 | 0.022 | <0.015 | <0.015 |
| | **SW8** | <0.015 | <0.015 | 0.014 | <0.005 | <0.025 | 0.007 | 0.065 | <0.025 | 0.304 | 0.018 | <0.015 |
| | **SW9** | <0.015 | <0.015 | 0.011 | 0.010 | <0.025 | 0.007 | 0.063 | <0.025 | 0.288 | 0.030 | <0.015 |
| | **SW10** | <0.015 | <0.015 | 0.009 | 0.009 | <0.025 | <0.005 | 0.085 | <0.025 | 0.204 | 0.042 | <0.015 |
| | **SW11** | <0.015 | <0.015 | 0.012 | 0.012 | <0.025 | 0.006 | 0.066 | <0.025 | 0.283 | <0.015 | <0.015 |
| Ground water | **GW1** | <0.015 | <0.015 | <0.005 | <0.005 | <0.025 | <0.005 | 0.035 | <0.025 | <0.005 | <0.015 | <0.015 |
| | **GW2** | <0.015 | <0.015 | <0.005 | <0.005 | <0.025 | <0.005 | 0.029 | <0.025 | <0.005 | <0.015 | <0.015 |
| | **GW3** | <0.015 | <0.015 | <0.005 | <0.005 | <0.025 | <0.005 | <0.015 | <0.025 | <0.005 | <0.015 | <0.015 |
| | **GW4** | <0.015 | <0.015 | <0.005 | <0.005 | <0.025 | <0.005 | <0.015 | <0.025 | <0.005 | <0.015 | <0.015 |
| | **GW5** | <0.015 | <0.015 | <0.005 | 0.015 | <0.025 | <0.005 | <0.015 | <0.025 | <0.005 | <0.015 | <0.015 |
| | **GW6** | <0.015 | <0.015 | <0.005 | <0.005 | <0.025 | <0.005 | <0.015 | <0.025 | <0.005 | <0.015 | <0.015 |
| **FD in SW (%)* | | 33 | 33 | 58 | 33 | 58 | 67 | 83 | 8 | 100 | 75 | 17 |
| **FD in all (%)* | | 21 | 21 | 32 | 26 | 37 | 42 | 53 | 5 | 63 | 47 | 11 |
| ***EQS (µg/L*)* | | – | 0.01 | – | – | – | – | 0.2 | – | 0.0083 | – | – |

*FD: frequency of detection in surface waters and all investigated waters
**EQS for annual average concentration for inland surface waters, or method detection limit in the case of Watch List pesticides (e.g., acetamiprid and imidacloprid)

Table 6. (continued)

| Sample type | Sample ID | MCPA | MCPP | Meta–laxyl | Pentachlorophenol | Propi–conazole | Pro–poxur | Simazine | Tebu–conazole | Terbutilazina–desethyl | Ter–butryn | Tri–clopyr |
|---|---|---|---|---|---|---|---|---|---|---|---|---|
| Surface water | **SW1** | <0.025 | 0.018 | 0.01 | <0.025 | <0.015 | <0.005 | <0.005 | 0.008 | <0.005 | 0.011 | <0.005 |
| | **SW2** | <0.025 | 0.070 | <0.005 | 0.201 | 0.037 | <0.005 | <0.005 | 0.019 | <0.005 | 0.099 | <0.005 |
| | **SW3** | <0.025 | 0.041 | <0.005 | 0.148 | 0.025 | <0.005 | <0.005 | 0.014 | <0.005 | 0.065 | <0.005 |
| | **SW4** | <0.025 | 0.150 | 0.012 | 0.378 | 0.031 | <0.005 | 0.013 | 0.027 | 0.009 | 0.262 | 0.02 |
| | **SW5** | <0.025 | 0.043 | <0.005 | 0.256 | 0.034 | 0.006 | <0.005 | 0.018 | <0.005 | 0.084 | <0.005 |
| | **SW6** | <0.025 | 0.078 | 0.006 | 0.246 | 0.032 | <0.005 | <0.005 | 0.025 | 0.005 | 0.107 | 0.009 |
| | **SW7** | <0.025 | <0.015 | <0.005 | <0.025 | <0.015 | <0.005 | <0.005 | <0.005 | <0.005 | 0.006 | <0.005 |
| | **SW8** | <0.025 | <0.015 | <0.005 | 0.070 | 0.015 | <0.005 | <0.005 | 0.010 | <0.005 | 0.071 | <0.005 |
| | **SW9** | <0.025 | <0.015 | <0.005 | 0.089 | <0.015 | <0.005 | <0.005 | 0.009 | <0.005 | 0.065 | <0.005 |
| | **SW10** | 0.049 | <0.015 | <0.005 | 0.061 | <0.015 | <0.005 | <0.005 | 0.007 | <0.005 | 0.050 | <0.005 |
| | **SW11** | <0.025 | <0.015 | <0.005 | 0.073 | <0.015 | <0.005 | <0.005 | 0.009 | <0.005 | 0.066 | <0.005 |
| Ground water | **GW1** | <0.025 | <0.015 | <0.005 | <0.025 | <0.015 | <0.005 | <0.005 | <0.005 | <0.005 | <0.005 | <0.005 |
| | **GW2** | <0.025 | <0.015 | <0.005 | <0.025 | <0.015 | <0.005 | <0.005 | <0.005 | <0.005 | 0.006 | <0.005 |
| | GW3 | <0.025 | <0.015 | <0.005 | <0.025 | <0.015 | <0.005 | <0.005 | <0.005 | <0.005 | <0.005 | <0.005 |
| | GW4 | <0.025 | <0.015 | <0.005 | <0.025 | <0.015 | <0.005 | <0.005 | <0.005 | <0.005 | <0.005 | <0.005 |
| | GW5 | <0.025 | <0.015 | <0.005 | <0.025 | <0.015 | <0.005 | <0.005 | <0.005 | <0.005 | <0.005 | <0.005 |
| | GW6 | <0.025 | <0.015 | <0.005 | <0.025 | <0.015 | <0.005 | <0.005 | <0.005 | <0.005 | <0.005 | <0.005 |
| **FD in SW (%)* | | 8 | 58 | 25 | 75 | 58 | 8 | 8 | 92 | 8 | 100 | 17 |
| **FD in all (%)* | | 5 | 32 | 16 | 47 | 32 | 5 | 5 | 53 | 5 | 68 | 11 |
| ***EQS (μg/L)* | | – | – | – | 0.4 | – | – | 1 | – | – | 0.065 | – |

*FD: frequency of detection in surface waters and all investigated waters

**EQS for annual average concentration for inland surface waters, or method detection limit in the case of Watch List pesticides (e.g., acetamiprid and imidacloprid)

Among the pesticides detected in surface waters, the herbicides terbutryn and imidacloprid were the most ubiquitous compounds as they were found in all samples in all sampling campaigns. Terbutryn use is currently banned in Europe and consequently in Spain. However, it was the most ubiquitous and abundant (up to 200 ng/g) pesticide recently detected in sediment samples of the lower Llobregat River basin (Barbieri et al., 2019), and therefore its presence could be attributed to its desorption from sediment where it may have accumulated during past applications. The application of imidacloprid is currently allowed only in greenhouses to protect tomato and zucchini crops. However, its physical-chemical characteristics (high water solubility 670 mg/L and low octanol-water partition coefficient (Log $K_{ow}$ = 0.57) favor its potential to reach waters. Diazinon (insecticide), diuron (herbicide), and carbendazim (fungicide) were also ubiquitous in surface waters, being present in all samples collected in winter and most of the samples investigated in summer. While diuron is still applied in Spain, although with more limitations than in the past, diazinon and carbendazim are not currently authorized for use as plant protection products in the area (information supplied by the Agrarian Park Consortium, the public entity in charge of the management and planning at the Agrarian Park of the Baix Llobregat). Therefore, these findings suggest that diuron is used upstream or for non-agricultural purposes, and that legacy pesticides can be also found in waters even after years of not being applied, as it is the case of diazinon.

Carbendazim, diuron, and imidacloprid, in addition to being ubiquitous in the area, were also among the most abundant pesticides found in the investigated surface waters, with an average concentration close to or above 0.1 μg/L in both sampling campaigns. Imidacloprid and the other neonicotinoid detected in the investigated waters, acetamiprid, were quantified at concentrations above the limit of detection (LOD) set in the Watch List for their analysis (EC, 2018), which corresponds to their lowest predicted no-effect concentrations (PNEC). Thus, the concentrations measured for these compounds could produce negative effects in exposed aquatic organisms. The same may be concluded for terbutryn and diuron in those locations where the concentrations measured surpassed their respective annual average EQS set for inland waters (Tables 5 and 6) (EC, 2013). Pentachlorophenol was also a relevant pesticide in terms of abundancy in both periods investigated, although its levels did not exceed the established annual average EQS (0.4 μg/L). Although its use is severely restricted because of its toxicity, persistence, and harmful effects on human health and the environment, this organochlorine persistent pesticide is still applied in the industry as a wood preservative.

No substantial seasonal differences were observed in the pattern of pesticides found in both surface and groundwater in the two periods monitored (winter *vs* summer) indicating a steady use or release of these compounds from soil and sediments where they may be accumulated throughout the year. Minor changes found were a larger contribution of imidacloprid, terbutryn, and pentachlorophenol in winter than in summer. The assignment of the currently used pesticides to a specific agronomic activity is rather challenging, due to the nature of the investigated area, where crops are highly diverse and cultivated land is highly fragmented (Figure S5). Imidacloprid, azoxystrobin, tebuconazole, and metalaxyl, all detected in both sampling campaigns, and acetamiprid, detected only in the summer campaign, are known to be applied throughout the year in the investigated area for the cultivation of artichoke, cucumber, tomato, and Brassica species, the main crops in the Agrarian Park (information supplied by the Agrarian Park Consortium, the public entity in charge of the management and planning at the Agrarian Park of the Baix Llobregat).

Despite the relatively high number of pesticides found in surface waters, only 7 were sporadically found in groundwaters (bentazone, carbendazim, diuron, pentachlorophenol, simazine, and terbutryn) and at levels below the maximum admissible concentration of 0.1 μg/L set in the legislation (EC, 2006). The occurrence of simazine, the only compound found in groundwater but not in surface water, whose use is currently banned, could be associated with its use in the past. Simazine has been found widespread in groundwaters of Catalonia in previous studies (Köck-Schulmeyer et al., 2014). The low concentrations of pesticides in the unconfined aquifer could be indicative of pollution sources other than current agricultural practices (e.g., aquifer recharge with surface water) or the capacity of the subsurface to naturally attenuate incoming pollutants (through sorption or degradation).

### 3.3. Origin of nitrate pollution

The aquifer system of the lower Llobregat River basin exhibits chronic pollution by both nitrate and ammonia, with a slow but continuous growing trend (Figure S6), as confirmed in this study.

In the winter survey, $\delta^{15}N\text{–}NO_3^-$ and $\delta^{18}O\text{–}NO_3^-$ were analyzed in all samples collected (Table 7) to discriminate between the organic and inorganic origin of nitrogen. In summer, in addition to $\delta^{15}N\text{–}NO_3^-$ and $\delta^{18}O\text{–}NO_3^-$, $\delta^{15}N\text{–}NH_4^{+,}$ and $\delta^{11}B$ were also determined in selected samples to distinguish between the different organic sources (Figure 4b and Table 8). Results are summarized in Figures 4 and 5, along with the theoretical boxes of the different sources as described in the literature (further details provided in section I and Figures S10 and S11).

Most of the samples investigated in winter (except SW5 and GW3) had $\delta^{18}O\text{–}NO_3^-$ slightly higher than the theoretical values expected for nitrification of $NH_4^+$ (Figure 4a), calculated following Mayer et al. (2001). However, the theoretical calculation of $\delta^{18}O\text{–}NO_3^-$ of nitrate derived from nitrification carries a degree of uncertainty (Snider et al., 2010), therefore the slightly high values $\delta^{18}O\text{–}NO_3^-$ could be attributed to either nitrification or denitrification (the latter would also increase $\delta^{15}N\text{–}NO_3^-$ values).

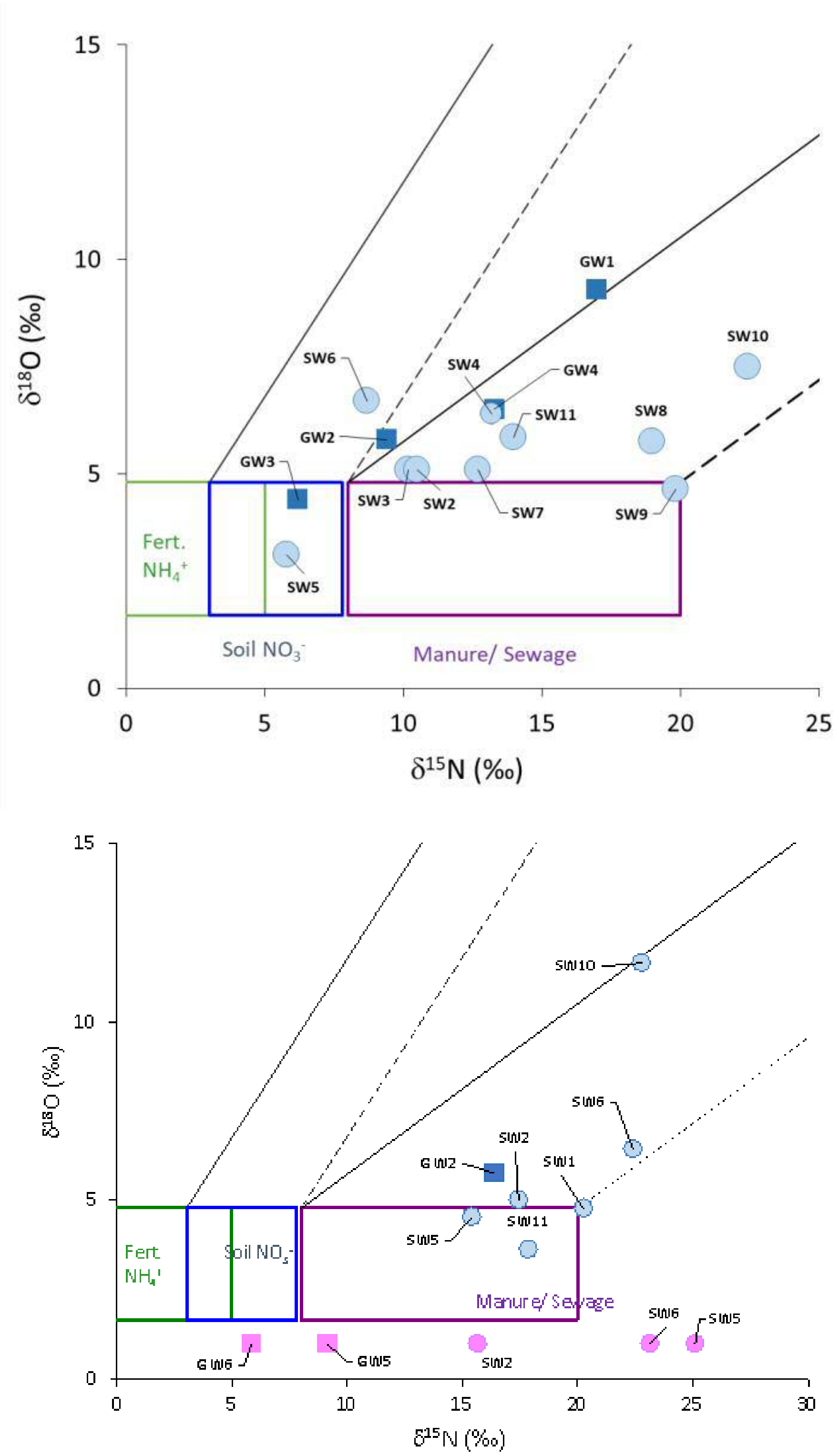

Figure 4. Graph $\delta^{15}$N–$NO_3^-$ *vs.* $\delta^{18}$O–$NO_3^-$ showing the same compositional boxes as in Figure S10 for the (a) winter and (b) summer survey campaigns. Blue circles correspond to surface water samples, and blue squares indicate groundwater samples. The $\delta^{15}$N–$NH_4^+$ has been represented (pink squares or circles) in the graph, using an arbitrary value of $\delta^{18}$O–$NO_3^-$ +1 ‰ for representation purposes.

Table 7. Results of the nitrate concentration and isotopic composition of the samples collected during the winter sampling campaign.

| | Sample | $NO_3^-$ [mg/L] | $\delta^{15}N\text{–}NO_3^-$ ± SD (‰) | $\delta^{18}O\text{–}NO_3^-$ ± SD (‰) |
|---|---|---|---|---|
| Surface water | **SW1** | 0.5 | – | – |
| | **SW2** | 15.8 | +10.5 ± 0.3 | +5.1 ± 0.2 |
| | **SW3** | 16.4 | +10.2 ± 0.3 | +5.1 ± 0.7 |
| | **SW4** | 8.4 | +13.2 ± 0.5 | +6.4 ± 0.2 |
| | **SW5** | 27.8 | +5.8 ± 0.3 | +3.1 ± 0.2 |
| | **SW6** | 13.4 | +8.7 ± 0.3 | +6.7 ± 0.2 |
| | **SW7** | 19.8 | +12.7 ± 0.3 | +5.1 ± 0.2 |
| | **SW8** | 26.6 | +19.0 ± 0.2 | +5.7 ± 0.3 |
| | **SW9** | 29.5 | +19.8 ± 0.6 | +4.6 ± 0.1 |
| | **SW10** | 22.2 | +22.4 ± 0.8 | +7.5 ± 0.2 |
| | **SW11** | 34.7 | +13.6 ± 0.9 | +5.9 ± 0.5 |
| Groundwater | **GW1** | 10.6 | +17.0 ± 0.9 | +9.3 ± 0.6 |
| | **GW2** | 40.5 | +9.4 ± 0.4 | +5.8 ± 0.5 |
| | **GW3** | 94.8 | +6.2 ± 0.9 | +4.4 ± 0.7 |
| | **GW4** | 13 | +13.3 ± 0.6 | +6.5 ± 0.5 |
| | **GW5** | <1.5 | n.d. | n.d. |
| | **GW6** | 6.0 | n.d. | n.d. |

n.d.= not determined because $NO_3$ was present at very low concentrations.
*SD: standard deviation of n=2 or 3 samples

Table 8. Results of the nitrate concentration and isotopic composition in the samples selected for analysis in the summer sampling campaign.

| | Sample | $NO_3^-$ [mg/L] | $NH_4^+$ [mg/L] | **δ**$^{15}$N-$NO_3^-$ ± SD **(‰)** | **δ**$^{18}$O-$NO_3^-$ ± SD **(‰)** | **δ**$^{15}$N-$NH_4^+$ ± SD **(‰)** | **δ**$^{11}$B ± SD **(‰)** |
|---|---|---|---|---|---|---|---|
| Surface water | SW1 | 5.9 | 0.2 | +16.9 ± 0.3 | +2.8 ± 0.4 | n.d. | – |
| | SW2 | 26.4 | 10 | +17.4 ± 0.3 | +5.0 ± 0.6 | +15.6 ± 0.4 | — |
| | SW5 | 23.2 | 12 | +15.4 ± 0.9 | +4.5 ± 0.7 | +25.1 ± 0.6 | – |
| | SW6 | 20.6 | 7.7 | +22.4 ± 0.5 | +6.4 ± 0.1 | +23.2 ± 0.2 | +6.9 ± 0.3 |
| | SW7 | 3.9 | <0.15 | +13.4 ± 0.8 | +11.9 ± 1.5 | n.d. | +3.6 ± 0.2 |
| | SW10 | 7.7 | <0.15 | +22.8 ± 0.4 | +11.6 ± 0.4 | n.d. | +10.5 ± 0.3 |
| | SW11 | 42.6 | 0.95 | +17.9 ± 0.0 | +3.6 ± 0.1 | n.d. | +9.2 ± 0.3 |
| Ground-water | GW2 | 19.3 | <0.15 | +16.4 ± 0.5 | +5.8 ± 0.8 | n.d. | – |
| | GW5 | <1.5 | 65 | n.d. | n.d. | +9.1 ± 0.1 | – |
| | GW6 | <12 | 37 | n.d. | n.d. | +5.9 ± 0.7 | – |

n.d.= not determined because $NO_3$ or $NH_4$ were present at very low concentrations.
*SD: standard deviation of n=2 or 3 samples

Only two surface water samples, namely SW5 and SW6 showed an isotopic signal that could be linked to chemical fertilizers (Figure 4a). Nevertheless, the $NH_4^+$ concentration observed in both samples (12 and 16 mg/L in SW5 and SW6, respectively) may suggest that nitrification is incomplete. In that case, the theoretical ranges for $NO_3^-$ sources cannot be used as they are calculated assuming that $NH_4^+$ nitrification is complete. The remaining surface water samples showed isotopic values compatible with an organic source, either manure or sewage. Among them, the most likely source of nitrate in SW2, SW3, SW4, and SW7 is linked to sewage or wastewater effluents, in line with the major land-use in these sampling locations (densely populated urban areas). The SW4 location receives the discharge of a WWTP and storm runoff, and this is reflected in its $\delta^{15}N$–$NO_3^-$ (+ 13.2 ‰) and $\delta^{18}O$–$NO_3^-$ (+ 6.4 ‰) values. The $\delta^{15}N$–$NO_3^-$ is in agreement with sewage values, while the $\delta^{18}O$–$NO_3^-$, slightly high, suggests either nitrification or a slight influence of denitrification that may have increased both the $\delta^{15}N$–$NO_3^-$ and the $\delta^{18}O$–$NO_3^-$ values. In contrast, in SW8, SW9, SW10, and SW11, located in an agricultural area, nitrate linked to livestock manure applied as organic fertilizers cannot be ruled out as a potential source of pollution. However, since these samples belong to a channel network that distributes treated wastewater (starting at SW8) for irrigation, the main source that contributes to these locations is also likely to be sewage.

Again in summer, the isotopic composition of $\delta^{15}N$–$NO_3^-$ vs. $\delta^{18}O$–$NO_3^-$ of selected surface water samples showed values in agreement with sewage/manure origin, even for SW5 and SW6. This was confirmed by $\delta^{15}N$–$NH_4^+$ (Figure 4b), and thus, an important contribution of chemical fertilizers, as suspected for SW5 and SW6 in the winter survey, can be discarded. According to the $\delta^{11}B$ results (Figure 5), all samples analyzed for which the organic origin of nitrate was uncertain, fitted with the theoretical field of sewage, discarding important contributions from livestock manure. In the case of SW7 (DWTP intake), the origin could be also associated with a mix of sewage and organic fertilizers, unfortunately, the low $NO_3^-$ concentration in summer did not allow confirmation of the winter $\delta^{15}N$–$NO_3^-$ vs. $\delta^{18}O$–$NO_3^-$ results. Thus, boron isotopes results were in agreement with the organic origin detected for most samples using N and O of dissolved nitrate.

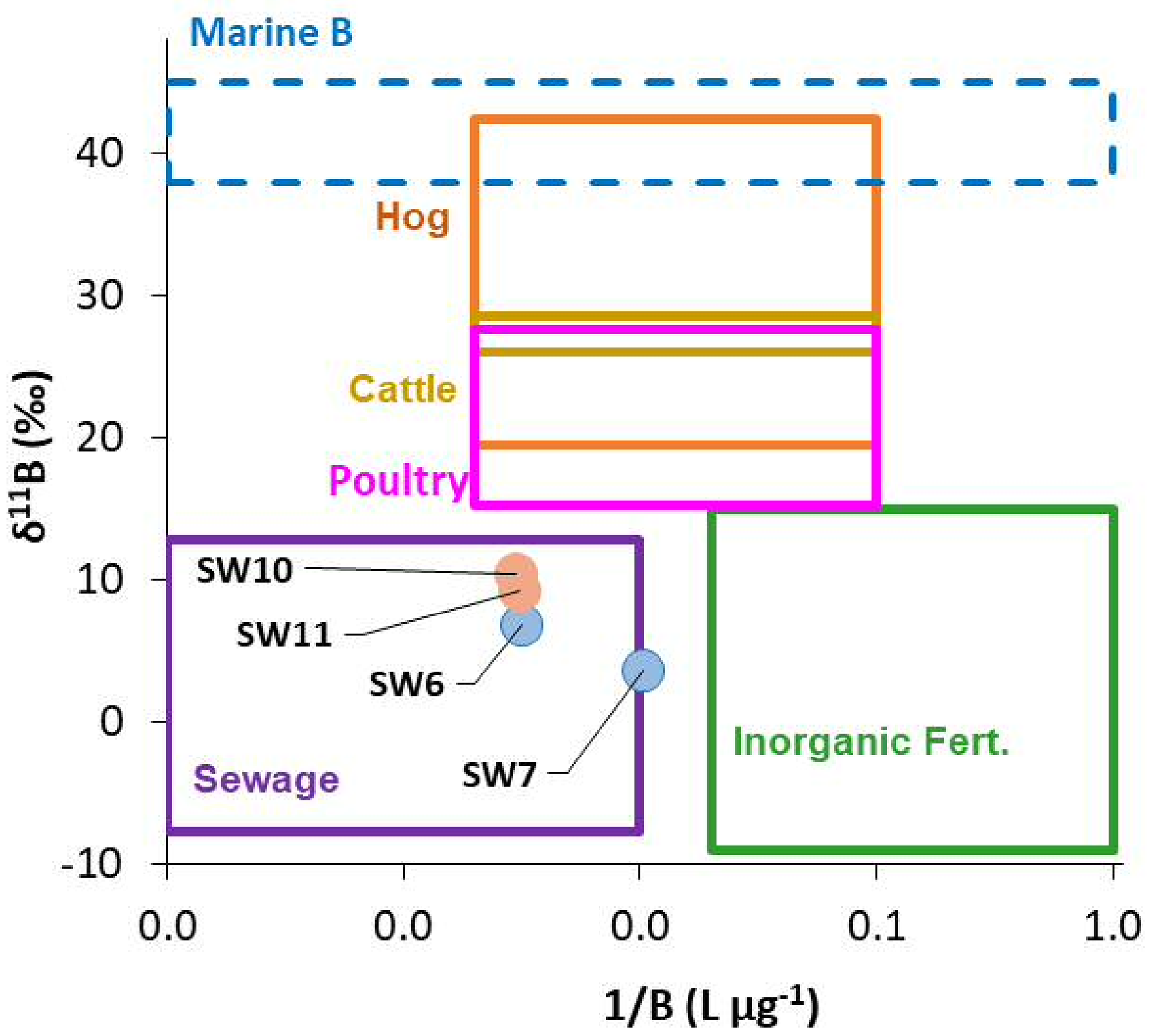

Figure 5. $\delta^{11}$B vs. 1/B diagram showing the compositional boxes described in (Widory et al., 2004; Widory et al., 2005) and including the results of the summer survey.

As for groundwaters, nitrate was detected in four (GW1, GW2, GW3, GW4) out of the six samples collected in winter. GW3 sample showed values in the area of soil nitrate. However, the same isotopic signature was also found in areas polluted exclusively with chemical fertilizers (Vitòria et al., 2005). This sample is located close to an area of intensive agriculture and had the highest nitrate concentration in the studied area (95 mg $NO_3^-$/L) discarding a natural origin of nitrate. Therefore, the main nitrate input affecting this sample could be linked to chemical fertilizers. The other three samples showed isotopic values affected by denitrification. Two of them, GW2 and GW1 lie in the area of uncertainty generated by denitrification where nitrate origin could be either related to chemical fertilizers, to an organic source, or a mix of both. The land uses in the surroundings of GW2 are both urban and agricultural, therefore nitrate could result from a mix of fertilizers and/or sewage. In contrast, as GW1 is located in an agricultural area, the most likely hypothesis is an origin of nitrate linked to fertilizers (either organic or inorganic). Nevertheless, since the application of sewage sludge as organic fertilizer is also a common practice in the agricultural fields under investigation this source cannot be completely ruled out as the source of nitrate pollution. Finally, GW4 lies in the area of denitrification from an organic source (either manure or sewage). The sample has the same isotopic composition of SW4 (with a high contribution of treated wastewater) (Figure 4a). However, these two samples are not close-by. Hence, for GW4, since the land use in the surroundings is a mixed agricultural industrial use, nitrate origin can be linked either to organic fertilizers and/or sewage.

In summer, the origin of N-species in those samples with non-detectable levels of $NO_3^-$(GW5 and GW6) was evaluated. These samples presented very high ammonium concentrations (65 mg $NH_4^+$/L in GW5 and 37 mg $NH_4^+$/L in GW6), despite the fact that ammonium is usually not found in groundwater because nitrification in the non-saturated zone takes place rather quickly. Such high ammonium levels in these parts of the aquifer could be associated with washing off of the accumulated organic matter in the subsoil or dissimilatory nitrate reduction to ammonium (DNRA) under reducing conditions. Since GW5 and GW6 belong to the surficial aquifer and the piezometric level of the aquifer varies considerably throughout the year (Figure S7) while ammonium concentrations remain fairly high year by year (monthly evolution data are not available), the first suggested source is not likely. In contrast, the high Fe and Mn concentrations found in these locations, and the low concentrations or absence of nitrate (Table S4), confirm the presence of a reducing environment, supporting the occurrence of DNRA processes. When nitrate is transformed into ammonium stable isotope fractionation processes occur. However, if the transformation is complete, the ammonium will have the same N isotopic composition of the initial nitrate. In the case of GW6, the source pollution is likely linked to chemical fertilizers (also supported by its location in an agricultural area). Furthermore, GW3, located upstream of GW6, showed nitrate concentrations close to 100 mg/L and a $\delta^{15}N-NO_3^-$ of +6.2 ‰. A total transformation of a similar nitrate concentration to ammonium will produce around 30 mg/L of ammonium with a similar isotopic composition, which fits with GW6 results. In the case of sample GW5, also located in an agricultural area, the isotopic signature observed could result from a mix of industrial fertilizers and sewage. This well is located nearby SW11, a sampling location where nitrate was derived from sewage.

*3.4. Identification of main pollution sources in the area*

The joint occurrence of N–nutrients and pesticides in surface waters is shown in Figures 6 and 7, and Figures S8–S9. Ammonium and nitrate exhibit a significant negative correlation in winter ($t(9) = -3.075$, *p–value* = 0.01363, r=–0.71) (Table 9), which could indicate that nitrate in surface water may be formed after nitrification of ammonium (Figure S8). In summer, this trend was not confirmed due to the low nitrate and ammonium concentrations observed in three locations SW1, SW7, and SW10. A significant positive and high correlation between ammonium concentration and total pesticides in the surface waters monitored in summer was observed (Figure 7) (Spearman's rank correlation, S=26, *p–value* < 0.001, r=0.88), which supports the hypothesis of a common source for both pollutants during this period. In this line, most of the ubiquitous ($n \geq 9$) and in some cases abundant pesticides detected in summer (i.e., diuron, tebuconazole, terbutryn, and pentachlorophenol) also showed a significant positive correlation with ammonia ($r > 0.71$) (Figure S9 and Table 10). In contrast, the correlations between ammonium and total or individual pesticide concentrations in winter were not significant in any case. Nitrate concentrations only correlated significantly with imidacloprid in summer ($t(7)=3.47$, *p–value*=0.0070, r=0.76) (Figure S9).

Considering this information and the occurrence of individual pesticides in the investigated waters, pesticide concentrations observed in surface water can be explained by two main sources. Given that N–species in the investigated surface water were associated with wastewater treatment plant discharges or sewage leakage, which rules out the relevance of agricultural activities in the nitrate and ammonium concentrations measured in surface water, the first source could be related to an urban use of these compounds. Furthermore, the facts that i) pesticide pollution patterns observed in treated wastewater or tributary rivers highly impacted with wastewater extended along with the irrigation channel networks that they feed, and ii) occasional pesticides (those with low detection frequency) were exclusively found within each irrigation channel network, support this hypothesis (e.g., terbuthylazine, metalaxyl, triclopyr, MCPP and 2,4-D in SW4 and SW6, or 2,4, MCPP, and acetamiprid in SW3 and SW5). Linking the pesticides observed in water to a specific crop in the area is not possible due to the high diversity of crops and rotation, and what is more important, the high fragmentation of the cultivated land (Figure S5).

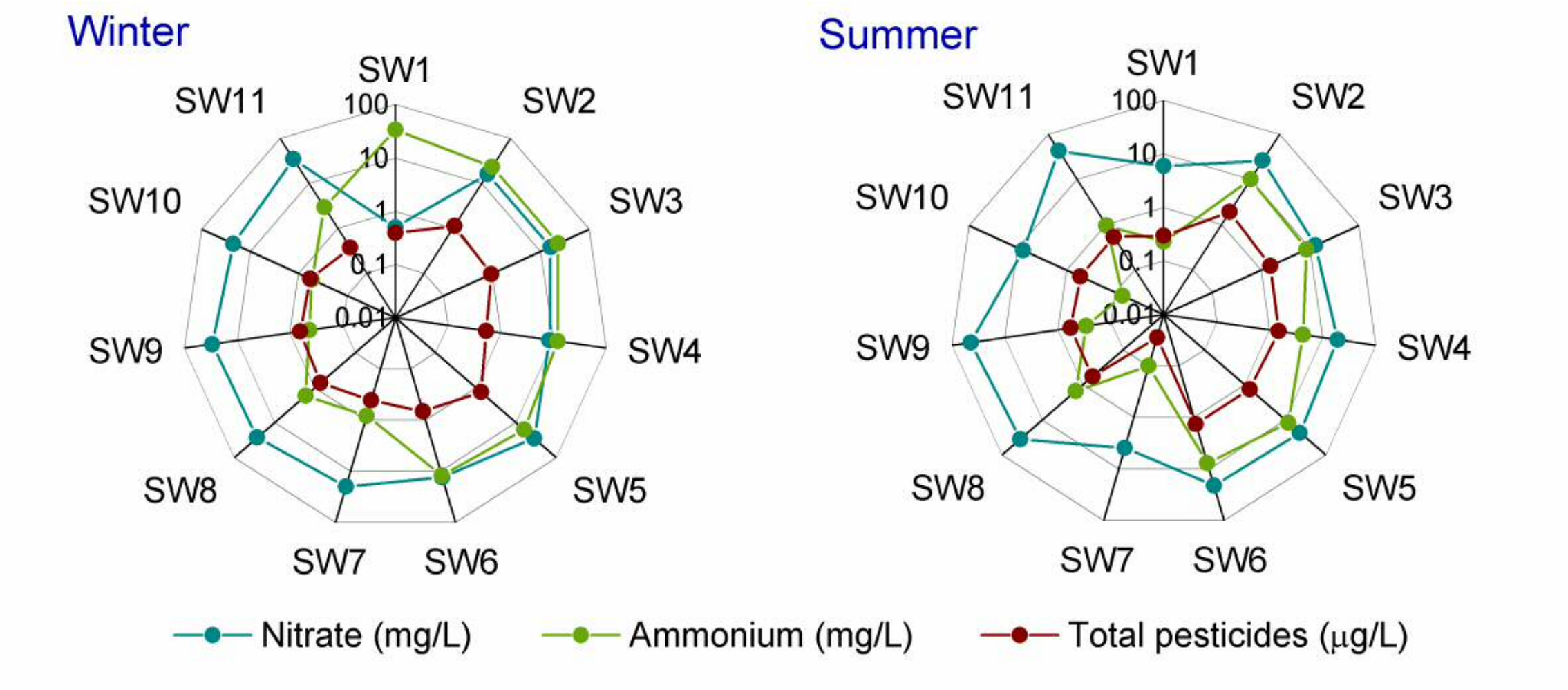

Figure 6. The occurrence of nitrates, ammonium, and total pesticides in the winter and summer sampling campaigns.

A second possible source of pesticides in the area could be attributed to desorption from soil and sediments of pesticides currently not in use. This kind of pesticides constitutes 16–59% of the total pesticide concentrations observed in all surface water samples except in SW7 in winter (92%). Higher contributions of legacy or banned pesticides were found in winter than in summer, which could be attributed to more frequent storm events, and hence higher river flow variations during winter and this may be the reason for not finding a significant correlation between ammonium and total pesticides in this period.

Based on the information provided by isotopes and pesticides, pollution from agricultural activities in surface waters is minor as compared to urban/industrial and legacy pollution sources.

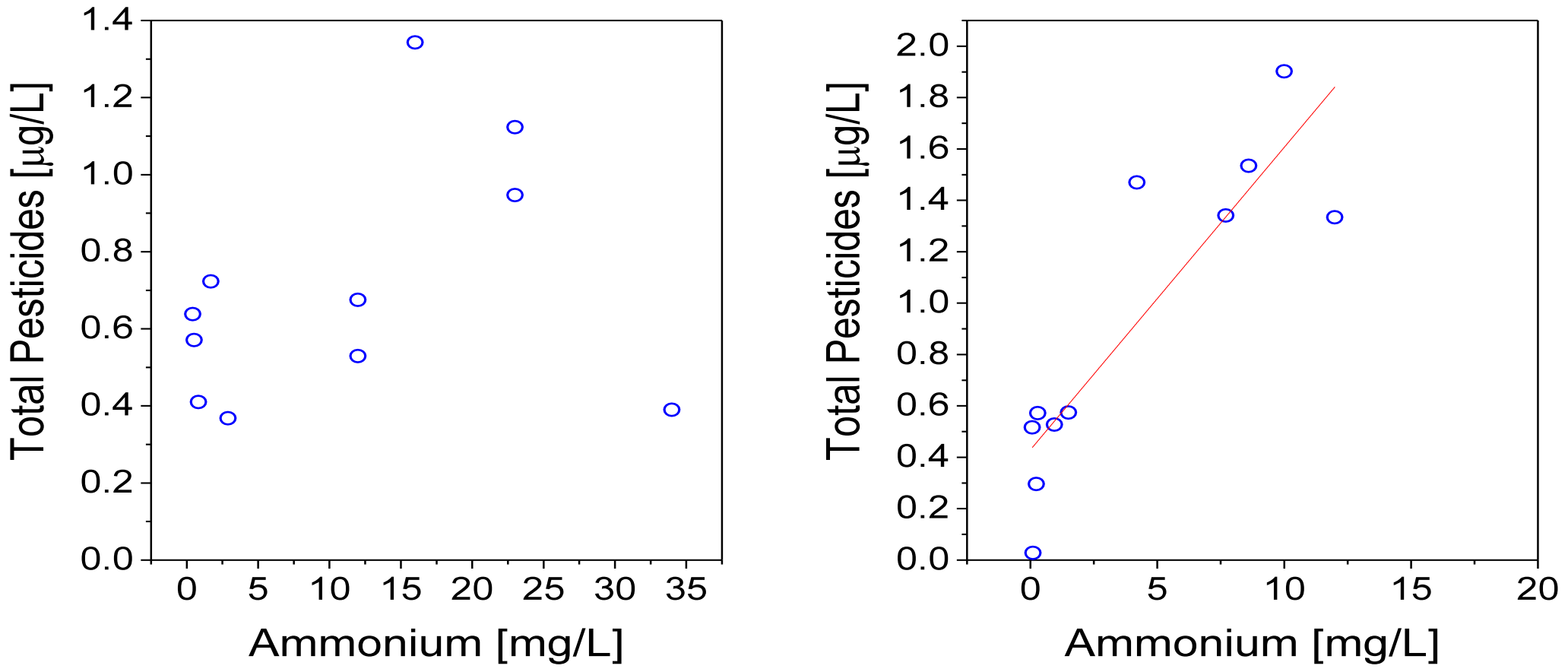

| Winter | Statistic | p–value | r | Summer | Statistic | p–value | r |
|---|---|---|---|---|---|---|---|
| $NH_4^+$ | t(9)=0.991 | 0.3478 | – | $NH_4^+$ | S=26 | 0.0007 | 0.88 |
| $NO_3^-$ | t(9)=0.471 | 0.6490 | – | $NO_3^-$ | t(9)=0.290 | 0.7783 | – |

Figure 7. Correlation between ammonium and total pesticides in surface waters in the winter and summer monitoring campaigns. Pearson's correlation test was used to evaluate the significance of the correlation between normally distributed variables (Shapiro–Wilk normality test), otherwise Spearman test was done. In both cases, the significance level was 0.05.

Table 9. Correlation between individual pesticide concentrations and ammonium or nitrate concentrations in winter. Pearson correlation test was conducted when variables distributed normally (after the Shapiro–Wilk normality test), otherwise, the Spearman correlation test was done.

| Winter | $NH_4^+$ | $NO_3^-$ |
|---|---|---|
| | *p-value* | *p-value* |
| $NH_4^+$ | 1 | |
| $NO_3^-$ | 0.01363 | 1 |
| Carbendazim | 0.2957 | 0.9682 |
| Diazinon | 0.0535 | 0.1534 |
| Diuron | 0.3251 | 0.9734 |
| Imidacloprid | 0.2080 | 0.120 |
| Tebuconazole | 0.9485 | 0.5722 |
| Terbutryn | 0.7083 | 0.0705 |

Table 10. Correlation between individual pesticide concentrations and ammonium or nitrate concentrations in summer. Pearson correlation test was conducted when variables distributed normally (after the Shapiro–Wilk normality test), otherwise, the Spearman correlation test was done.

| Summer | $NH_4^+$ | $NO_3^-$ |
|---|---|---|
| | *p-value* | *p-value* |
| $NH_4^+$ | 1 | – |
| $NO_3^-$ | 0.2606 | 1 |
| Diuron | 0.0041 | 0.7892 |
| Imidacloprid | 0.7757 | 0.0070 |
| Pentachlorophenol | 0.0369 | 0.3628 |
| Tebuconazole | 0.0065 | 0.7395 |
| Terbutryn | 0.0076 | 0.1454 |

As for groundwater, the fact that pesticide concentrations were very low, likely due to their natural attenuation in the subsurface, complicates the assessment of the main pollution sources. Moreover, the high variability of the piezometric level of the aquifer in the area throughout the year also complicates the scenario. The analysis of stable isotopes in nitrate and ammonium and land uses in the area pointed out the agricultural activity as the main pollution source at least in GW1, GW3, GW4, and GW6, while urban and agricultural sources would mainly affect GW2 and GW5. Pesticides found in the unconfined aquifers were the same as those found in surface waters, but they were only detected occasionally in few locations (GW1, GW2, GW5, and GW6), and thus leakage of the irrigation infrastructure may be the main source of pesticides into the aquifer, rather than diffuse pollution caused by agricultural activities. The only minor exception might be the triazine herbicide simazine found in winter in GW2 and absent in surface waters, which is attributable to its use in the past and its desorption from the soil where it may be accumulated. Different pesticides were found in reductive environments (bentazone in GW5 and pentachlorophenol in GW6) as compared to oxidative parts of the aquifer (diuron, carbendazim, and terbutryn in GW1 and simazine and diuron in GW2). As expected, pesticide pollution is absent in the deep aquifer (GW4), even though nitrate and ammonium were found.

## 4. Conclusions

The protection of water resources requires the implementation of measures to control pollution sources. This study presents a combined approach that evaluates pesticide occurrence and the origin of N-species in strategic locations of a multi-stressed watershed to assess the main pollution sources that deteriorate water quality. Stable isotopes were used to evaluate the origin of nitrate and ammonium concentrations measured in groundwater and surface water. This, in combination with observed patterns of pesticide occurrence, local land uses, and hydrodynamics, helped discriminate the origin of the pollution observed. The approach presented here is an example of the type of the multiple lines of evidence approach to investigative monitoring envisioned in the WFD.

Its application to the Lower Llobregat River basin revealed urban/industrial activities as the main pressure on the quality of surface water resources, and to a large extent also on groundwater resources, although agriculture may also play an important role, mainly in terms of nitrate and ammonium pollution of groundwater. Pesticide pollution in groundwater was much lower than in surface water (0.2 *vs* 1.9 µg/L), likely due to natural attenuation of contaminants in the subsurface, and may have its origin on surface waters. Nitrate in groundwater may result partially or totally from the nitrification of ammonium from an organic source (sewage or manure), except in GW3 where it is originated by the use of inorganic fertilizers. Several wells showed ammonium pollution, probably generated by nitrate reduction within the aquifer, but a more detailed sampling would be required to confirm this. In the analyzed samples, no significant attenuation processes have been observed for nitrate pollution. Results of the boron isotopes suggested that the organic origin of nitrate is linked to sewage, but this conclusion cannot be extrapolated to the whole area due to the limited number of samples analyzed.

The investigative monitoring conducted in this study allows the identification of the most relevant pollution sources at local level, which is needed to develop targeted and effective mitigation strategies to reduce pollutants and protect water resources.

## Acknowledgments

**This work has received funding from the EU's** Horizon 2020 Research and Innovation Programme through the WaterProtect project (grant agreement No. 727450), the Spanish Ministry of Science and Innovation (Project CEX2018–000794–S), and the Generalitat de Catalunya (Consolidated Research Group 2017 SGR 01404–Water and Soil Quality Unit).

SUPPLEMENTARY MATERIAL

# Investigative monitoring of pesticide and nitrogen pollution sources in a complex multi-stressed catchment: the Lower Llobregat River basin case study (Barcelona, Spain)

Cristina Postigo[1*], Antoni Ginebreda[1*], Maria Vittoria Barbieri[1], Damià Barceló[1,2], Jordi Martin[3], Agustina de la Cal[3], Maria Rosa Boleda[3], Neus Otero[4,5], Raul Carrey[4], Vinyet Solá[6], Enric Queralt[6], Elena Isla[7], Anna Casanovas[7], Gemma Frances[7], Miren López de Alda[1]

[1] *Department of Environmental Chemistry, Institute of Environmental Assessment and Water Research (IDAEA-CSIC), Carrer de Jordi Girona 18-26, 08034, Barcelona, Spain*

[2] *Catalan Institute for Water Research (ICRA), Emili Grahit, 101, Edifici H2O, Parc Científic i Tecnològic de la Universitat de Girona, 17003 Girona, Spain.*

[3] *Aigües de Barcelona, Empresa Metropolitana de Gestió del Cicle Integral de l'Aigua, S.A. Carrer de General Batet 1-7, 08028 Barcelona, Spain.*

[4] *Grup MAiMA, SGR Mineralogia Aplicada, Geoquímica i Geomicrobiologia, Departament de Mineralogia, Petrologia i Geologia Aplicada, Facultat de Ciències de la Terra, Universitat de Barcelona (UB), Carrer de Martí i Franquès s/n, 08028 Barcelona, Spain, i Institut de Recerca* ***de l'Aigua (IdRA)****, UB.*

[5] *Serra Húnter Fellowship, Generalitat de Catalunya, Spain.*

[6] ***Comunitat d'Usuaris d'Aigües de la Vall Baixa i del Delta del Llobregat (CUADLL)****, Carrer de Pau Casals 14-16, local, 08820, El Prat de Llobregat, Spain.*

[7] *Parc Agrari del Baix Llobregat, Can Comas, Camí de la Rivera, s/n, 08820 El Prat de Llobregat, Spain;*

*Corresponding authors:

Cristina Postigo (0000-0002-7344-7044) cprgam@cid.csic.es
Antoni Ginebreda (0000-0003-4714-2850) agmgam@cid.csic.es
Institute of Environmental Assessment and Water Research (IDAEA-CSIC)
Department of Environmental Chemistry
C/ Jordi Girona 18-26, 08034 Barcelona, Spain.
Tel: +34-934-006-100, Fax: +34-932-045-904

*Number of pages: 19*

*Number of figures: 11*

*List of Figures:*

*Other contents*

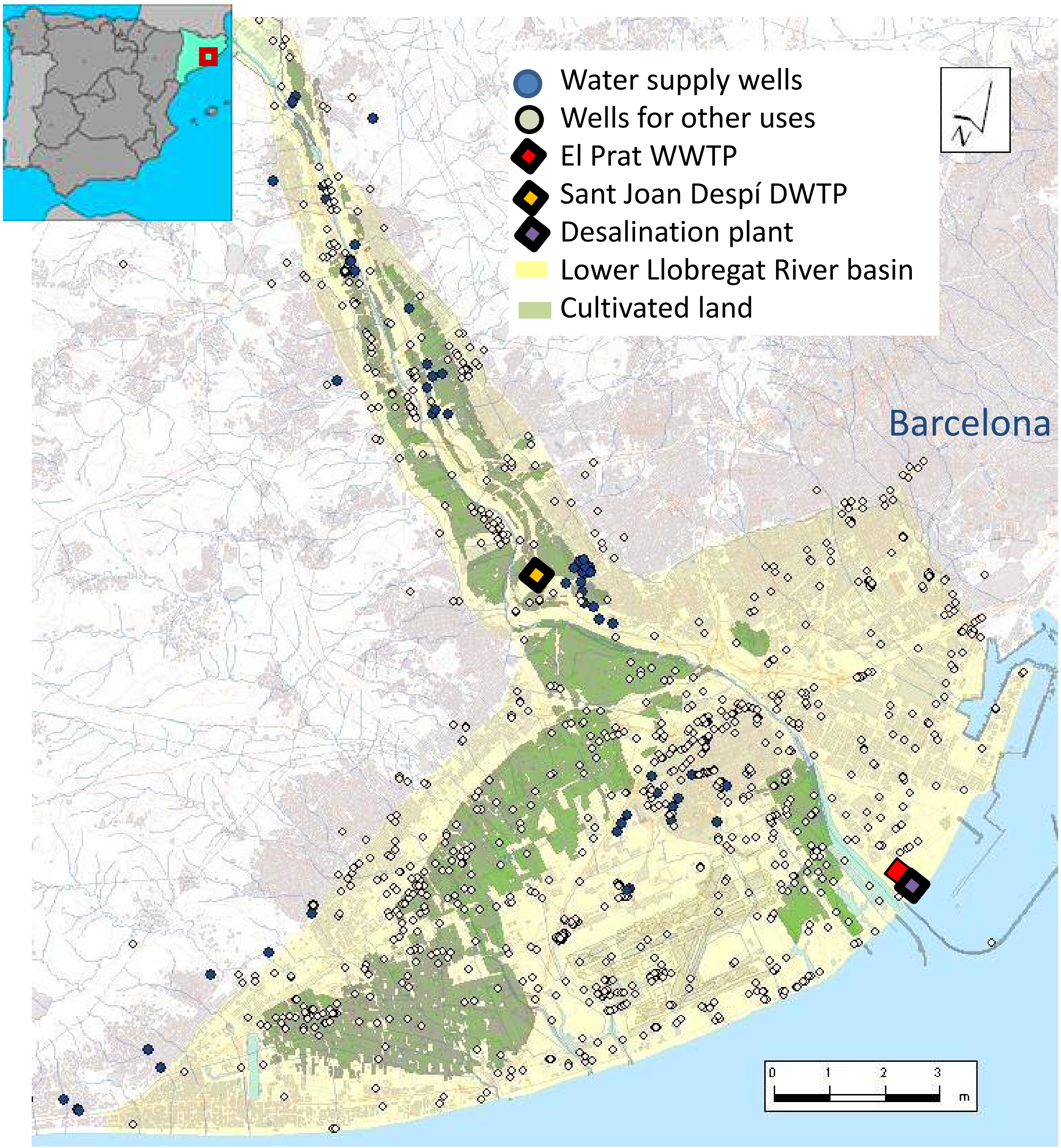

Figure S1. Map of the lower Llobregat River basin showing the location of water extraction wells, the main WWTP and DWTP, and the desalination plant.

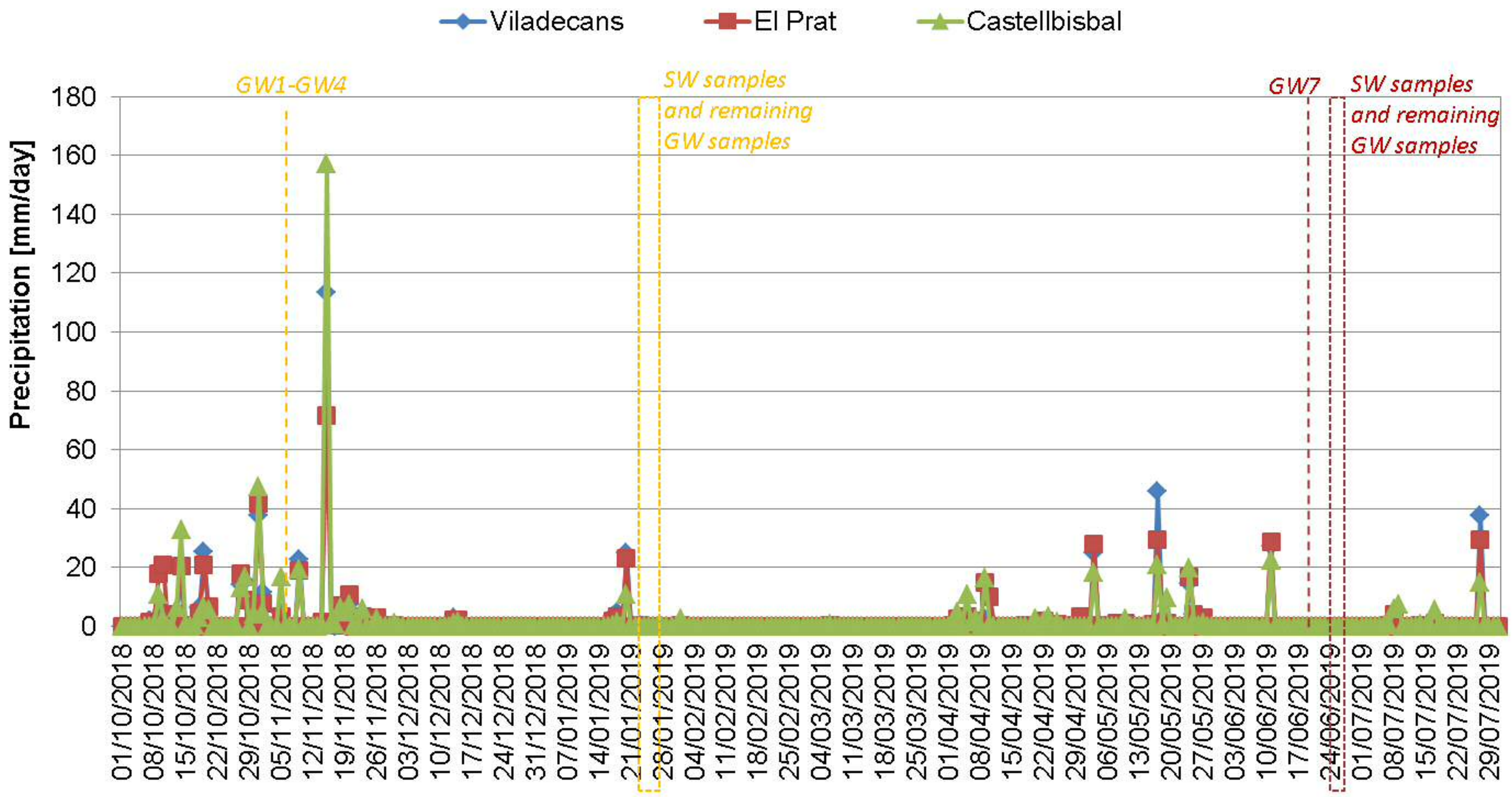

Figure S2. Precipitation (mm/day) registered in three stations located in the lower Llobregat River basin during the sampling periods (orange: winter campaign, red: summer campaign).

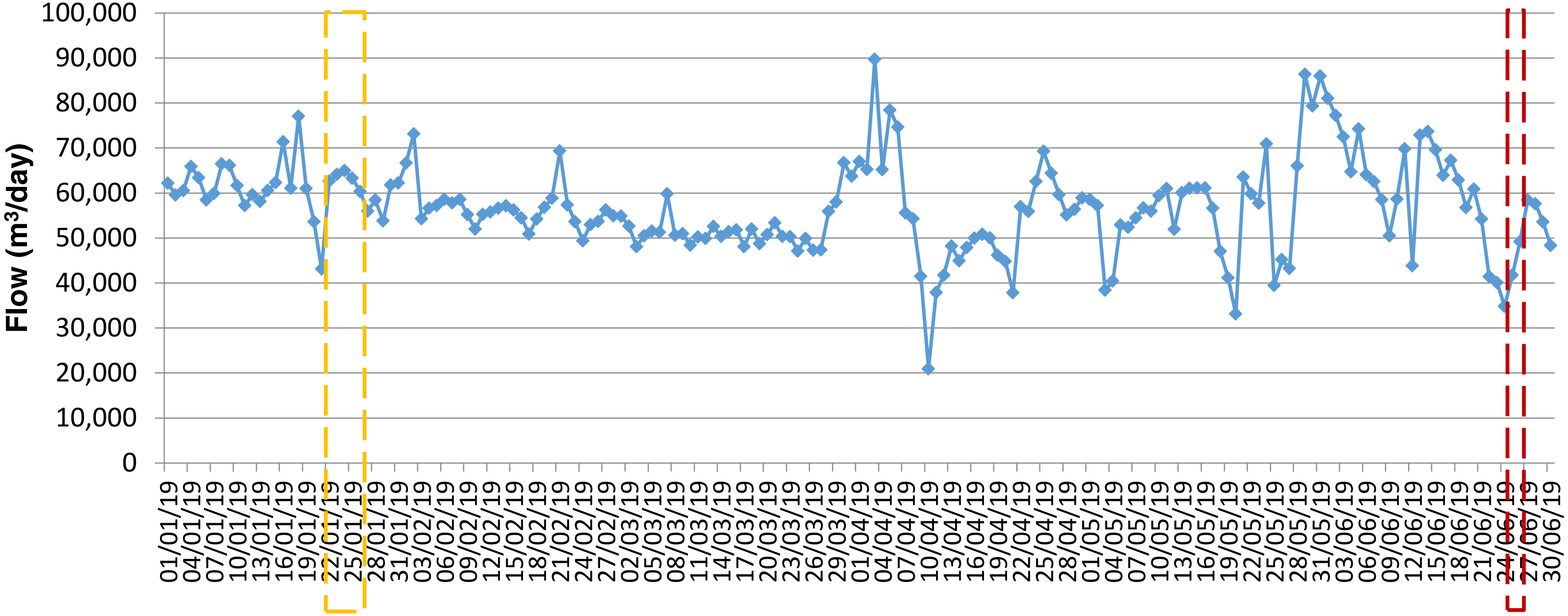

Figure S3. Daily average flow ($m^3/s$) in the main Llobregat River (St Vicenç dels Horts gauging station) during the sampling periods (orange: winter campaign, red: summer campaign).

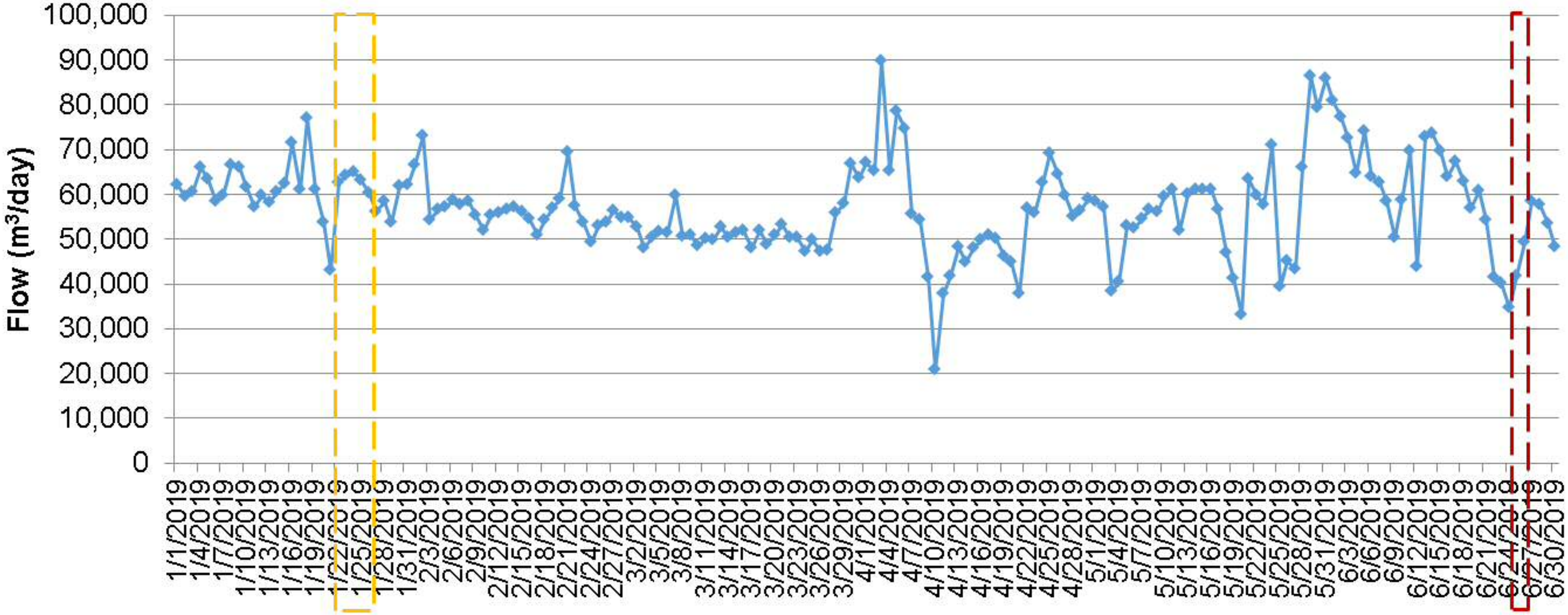

Figure S4. Daily average flow ($m^3$/day) diverted from the Anoia tributary and the Riera Creek into the Infanta Channel irrigation network (SW3) (St Vicenç dels Horts gauging station) during the sampling periods (orange: winter, red: summer).

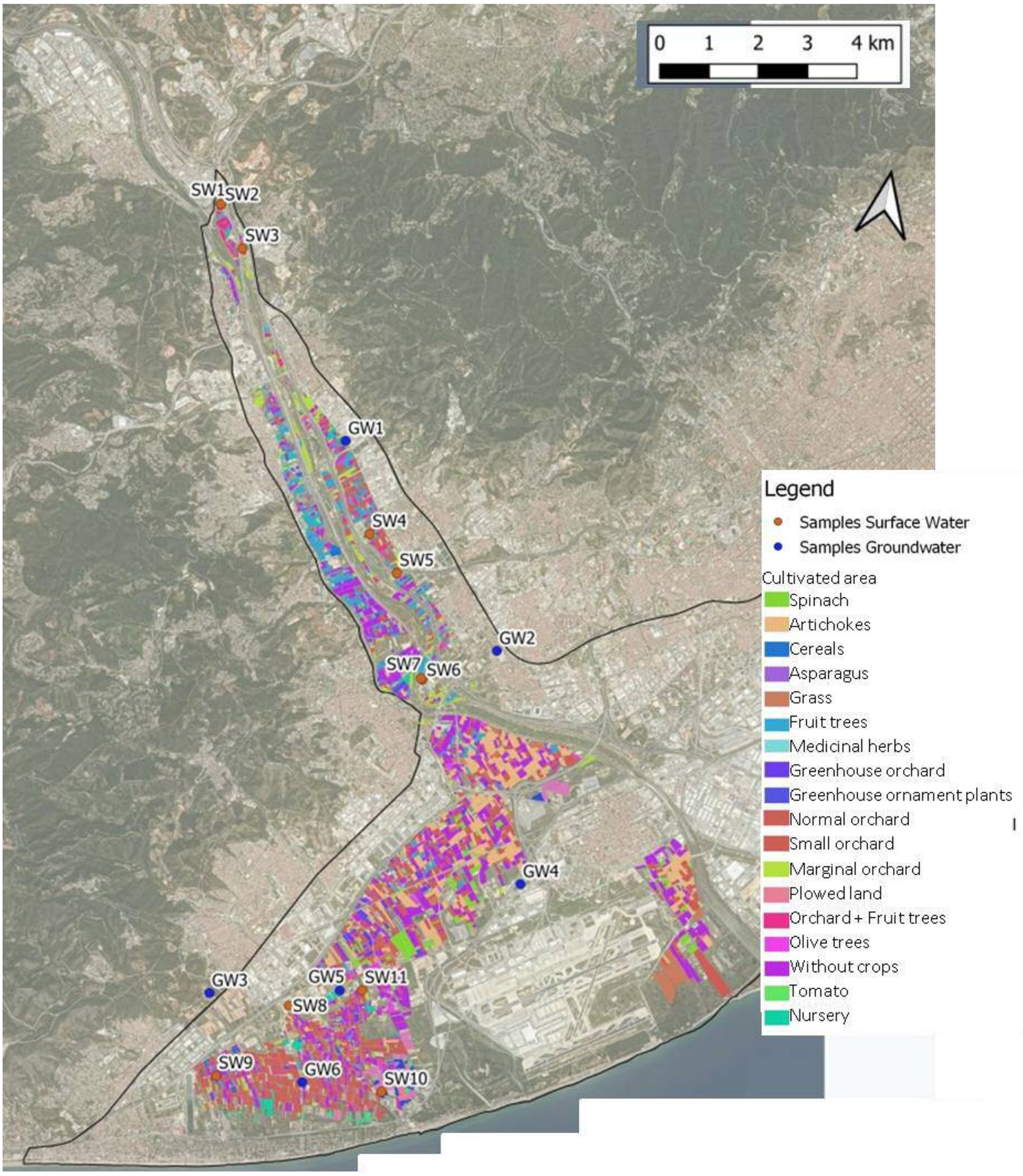

Figure S5. Map of the lower Llobregat River basin showing the high diversity of crops and fragmentation of cultivated land in the area.

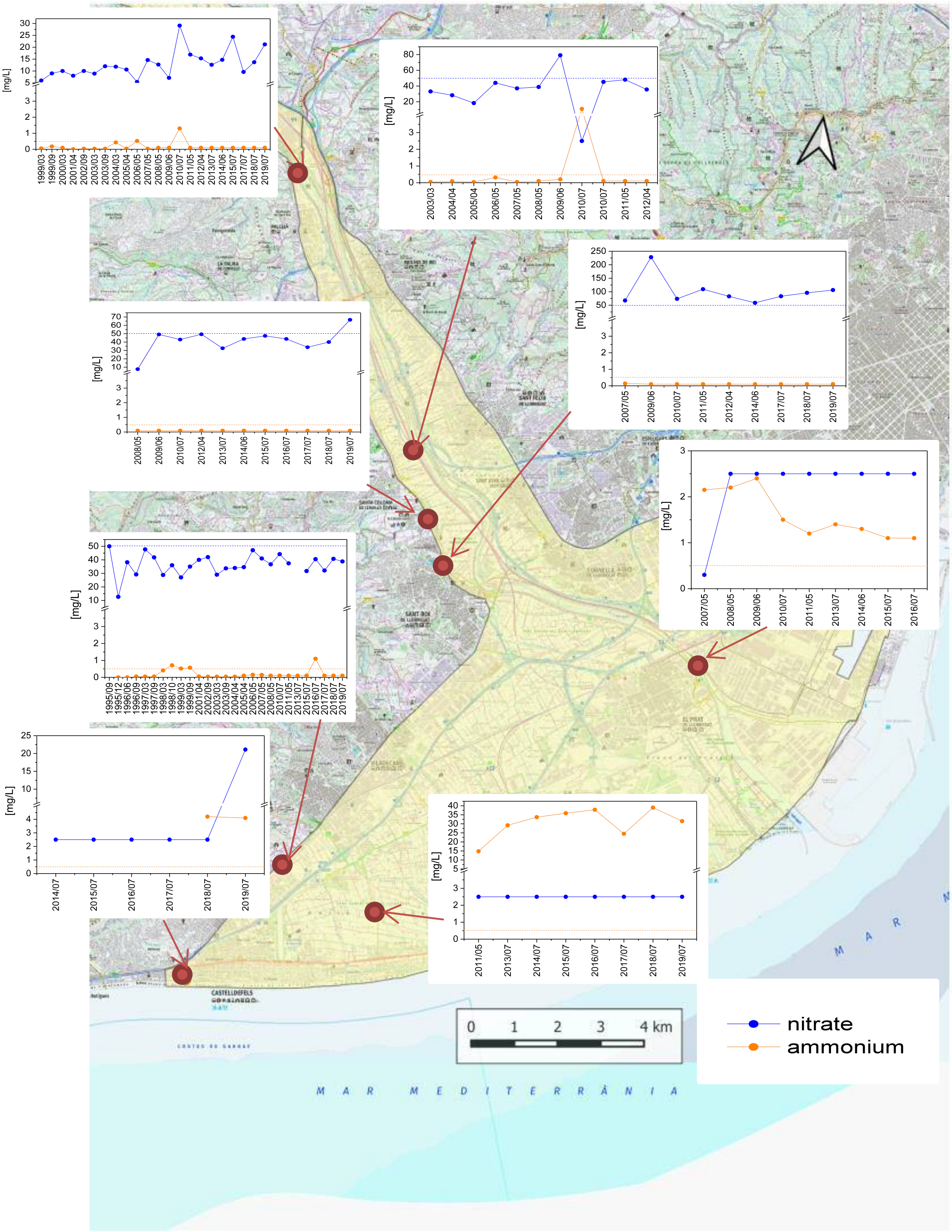

Figure S6. Historic pollution of nitrate and ammonium in the lower Llobregat River aquifer system. Water quality data obtained from the Catalan Water Agency (http://aca-web.gencat.cat/sdim21/).

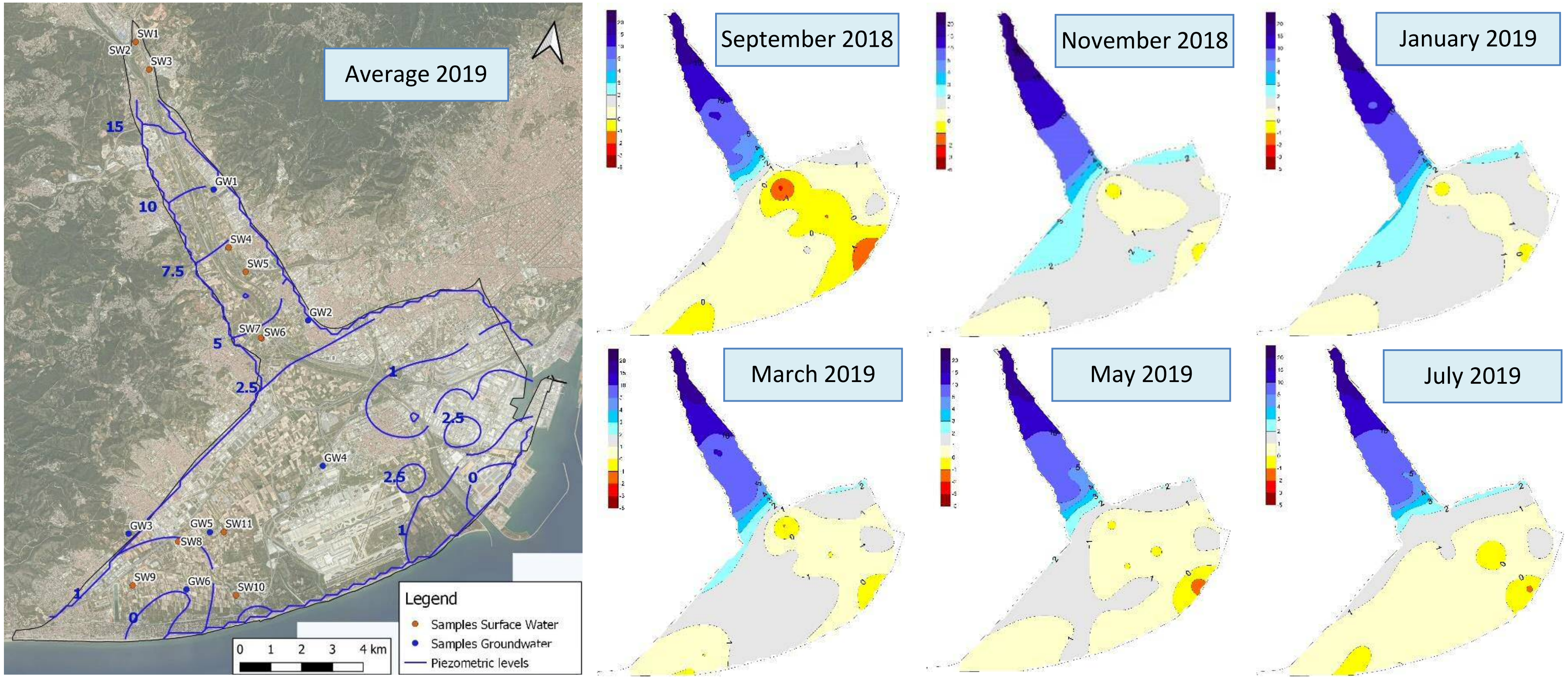

Figure S7. Piezometric levels (meters above sea level, masl) in the investigated area during the sampling period.

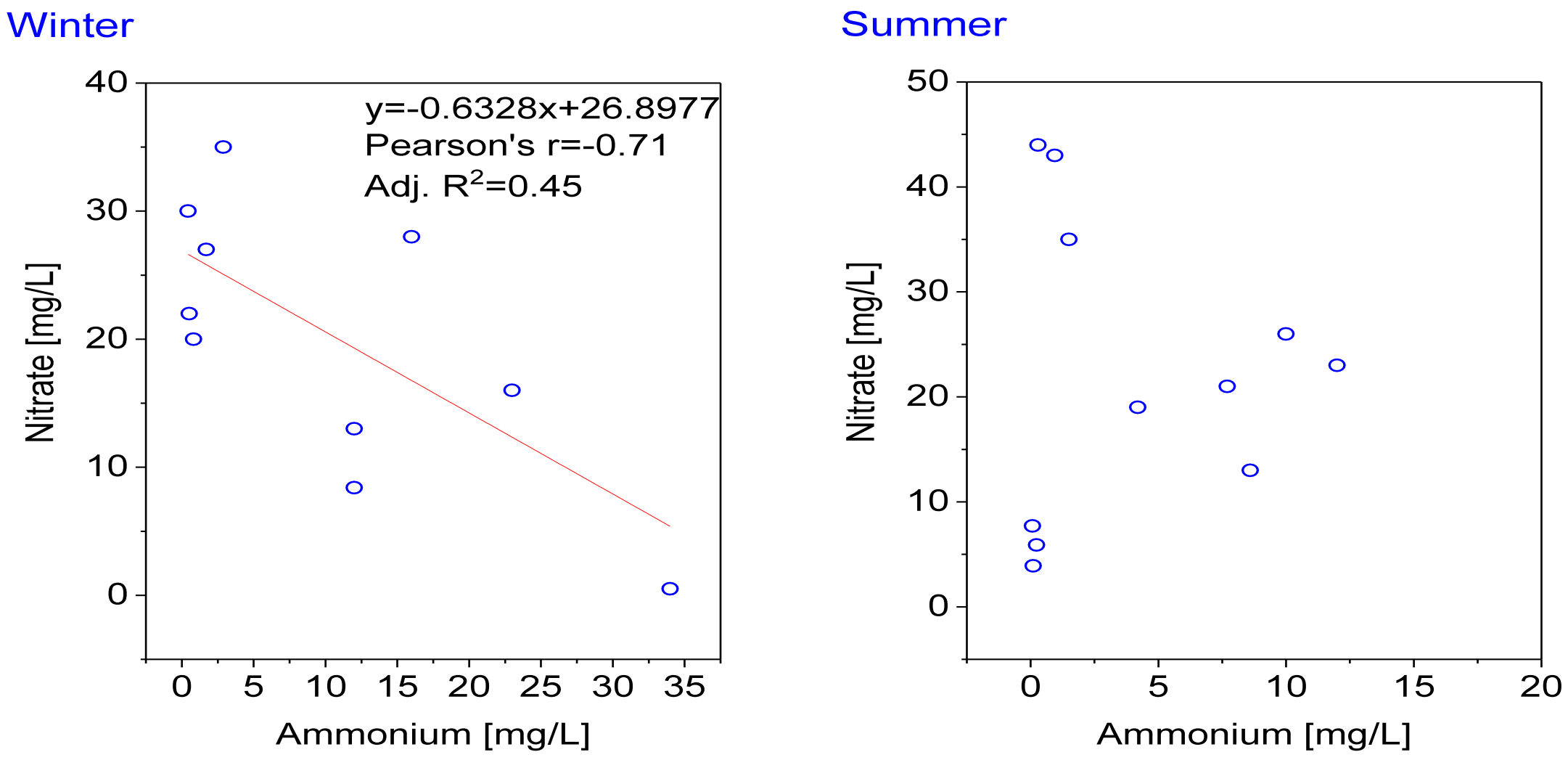

Figure S8. Correlation between ammonium and nitrate concentrations.

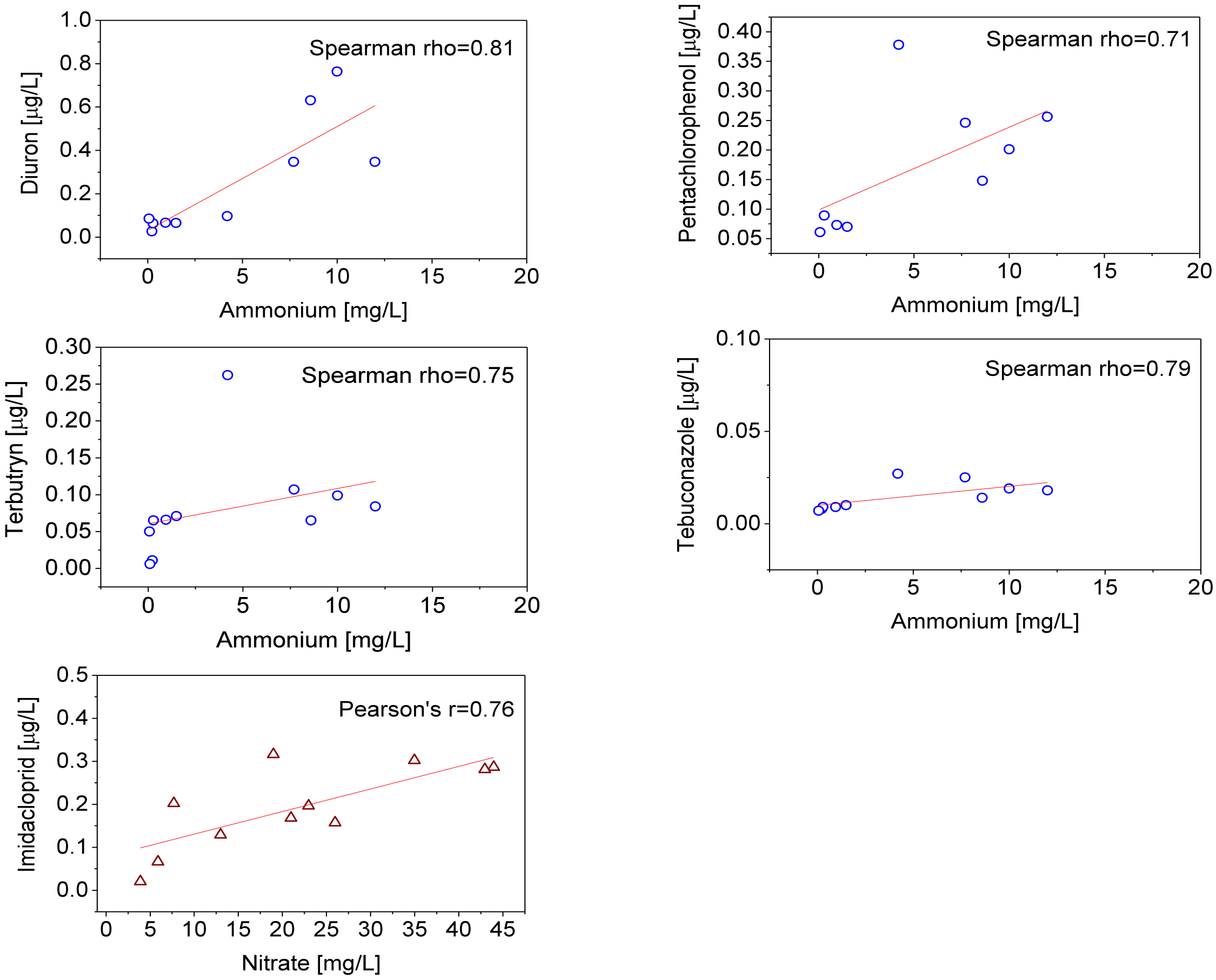

Figure S9. Significant correlations between ammonium or nitrate concentrations and individual pesticide concentrations. All corresponded to the summer period.

# I. Interpretation of N, O, and B stable isotopic data: an overview

The isotopic composition of certain chemical compounds allows discriminating their origin (Aravena and Robertson, 1998; Clark, 1997). Moreover, depending on the processes undergone by the compound, the isotopic composition may vary and therefore also reveals the occurrence of processes, *e.g.*, the existence of natural attenuation (denitrification in the case of nitrate pollution).

### ***$^{15}N$ and $^{18}O$ in nitrogen-containing compounds***

Nitrogen pollution can be related to sanitation network leakage, urban landfill leachate, and the use of synthetic fertilizers and/or manure. The analysis of the $\delta^{15}N$–$NO_3^-$ and $\delta^{18}O$–$NO_3^-$ of dissolved nitrate and $\delta^{15}N$–$NH_4^+$ of dissolved ammonium allows identifying to some extent the origin of these nitrogen compounds. Table S11 summarizes the expected $\delta$ values of the stable isotopes of N and O in nitrate and ammonium according to their origin.

Table S1. Stable isotope composition of N–compounds according to their origin (Vitòria et al., 2004).

| Origin | $\delta^{15}N$-$NH_4^+$ | $\delta^{15}N$-$NO_3^-$ | $\delta^{18}O$-$NO_3^-$ |
|---|---|---|---|
| Fertilizers (inorganic) | −5 ‰ –+ 5 ‰ | −5 ‰ –+ 5 ‰ | + 23 ‰. |
| Animal manure and urban wastewater (organic) | + 8 ‰ –+ 20 ‰ | + 8 ‰ –+ 20 ‰ | 0 –+ 7 ‰ |
| Soil organic matter (organic) | | + 3 ‰ –+ 9 ‰ | |
| Nitrification | | | +1.7 ‰ –+ 4.8 ‰ |

As shown in Table S11, nitrate from organic waste effluents present a $\delta^{18}O$–$NO_3^-$ lower than that of fertilizers, since it is generated by nitrification of ammonium or organic–bonded compounds. During this process, the oxygen incorporated in the nitrate molecule is derived from water oxygen and atmospheric oxygen (Mayer et al., 2001), and this allows discriminating between fertilizers, and livestock/urban sanitation networks (see diagram $\delta^{15}$N–$NO_3^-$ vs. $\delta^{18}$O–$NO_3^-$ in Figure S8).

Nitrate derived from nitrification, either from inorganic ammonium fertilizers, from soil organic nitrogen, manure, or septic tanks and sewage networks, will have a characteristic $\delta^{18}$O–$NO_3^-$ that will be based on $\delta^{18}$O–$H_2O$ of local groundwater and $\delta^{18}$O–$O_2$ of oxygen dissolved in water (Mayer et al., 2001):

$$\delta^{18}O_{NO3} = 1/3\ \delta^{18}O_{O_2} + 2/3\ \delta^{18}O_{H_2O} \quad \text{(Eq. 1)}$$

where the value of $\delta^{18}$O–$O_2$ is + 23 ‰ and the value of $\delta^{18}$O–$H_2O$ is variable and depends on the study area. In Figure S8, the compositional boxes calculated with the minimum and maximum values of $\delta^{18}$O–$H_2O$ of groundwater of Mediterranean influence (Araguas–araguas, 2005) are presented.

Furthermore, when interpreting isotopic data of N compounds is also necessary to consider that the isotopic composition can be affected by various physical, chemical, and biological processes, such as volatilization, denitrification, mixing, etc.

Ammonium ($NH_4^+$), the main constituent of inorganic fertilizers, livestock manure, and wastewater, is affected by volatilization processes. Volatilization will produce an increase in $\delta^{15}$N–$NH_4^+$ (Mariotti et al., 1981). This process is represented in Figure S8 with a horizontal arrow.

One of the processes that can significantly modify the N and O isotopic composition of dissolved nitrate is denitrification (reduction of $NO_3^-$ to $N_2$), natural attenuation of nitrate pollution. This reaction occurs through different intermediate stages.

$$NO_3^-(aq) \rightarrow NO_2^-(aq) \rightarrow N_XO\ (g) \rightarrow N_2\ (g) \quad \text{(Eq. 2)}$$

This redox process acts under anaerobic conditions where bacteria obtain energy from the reduction of $NO_3^-$ and the oxidation of organic matter or inorganic compounds such as sulfides (Rivett et al., 2008). During $NO_3^-$ reduction, there is an isotopic fractionation that increases the isotopic composition, both in $\delta^{15}N-NO_3^-$ and $\delta^{18}O-NO_3^-$ of the remaining nitrate with a characteristic slope (Böttcher et al., 1990; Fukada et al., 2003). Therefore, the use of stable isotopes is a very useful tool to determine the existence of denitrification processes since nitrate affected by denitrification will present an isotopic composition that will be projected on the theoretical field framed between these two bibliographic slopes. The variability of both slopes generates a zone of overlap or uncertainty where the origin of nitrate cannot be univocally assigned (Fig. S8).

Finally, in soils with high microbial activity, nitrate fertilizers can be recycled in the soil in a process abbreviated as MIT (Mineralization – Immobilization – Turnover), (Mengis et al., 2001). During this process, the isotopic composition of the N is approximately constant, but the $\delta^{18}O_{NO3}^-$ loses its characteristic isotopic signal of + 23 ‰ and will have the same $\delta^{18}O_{NO3}^-$ that nitrified ammonium-based fertilizers.

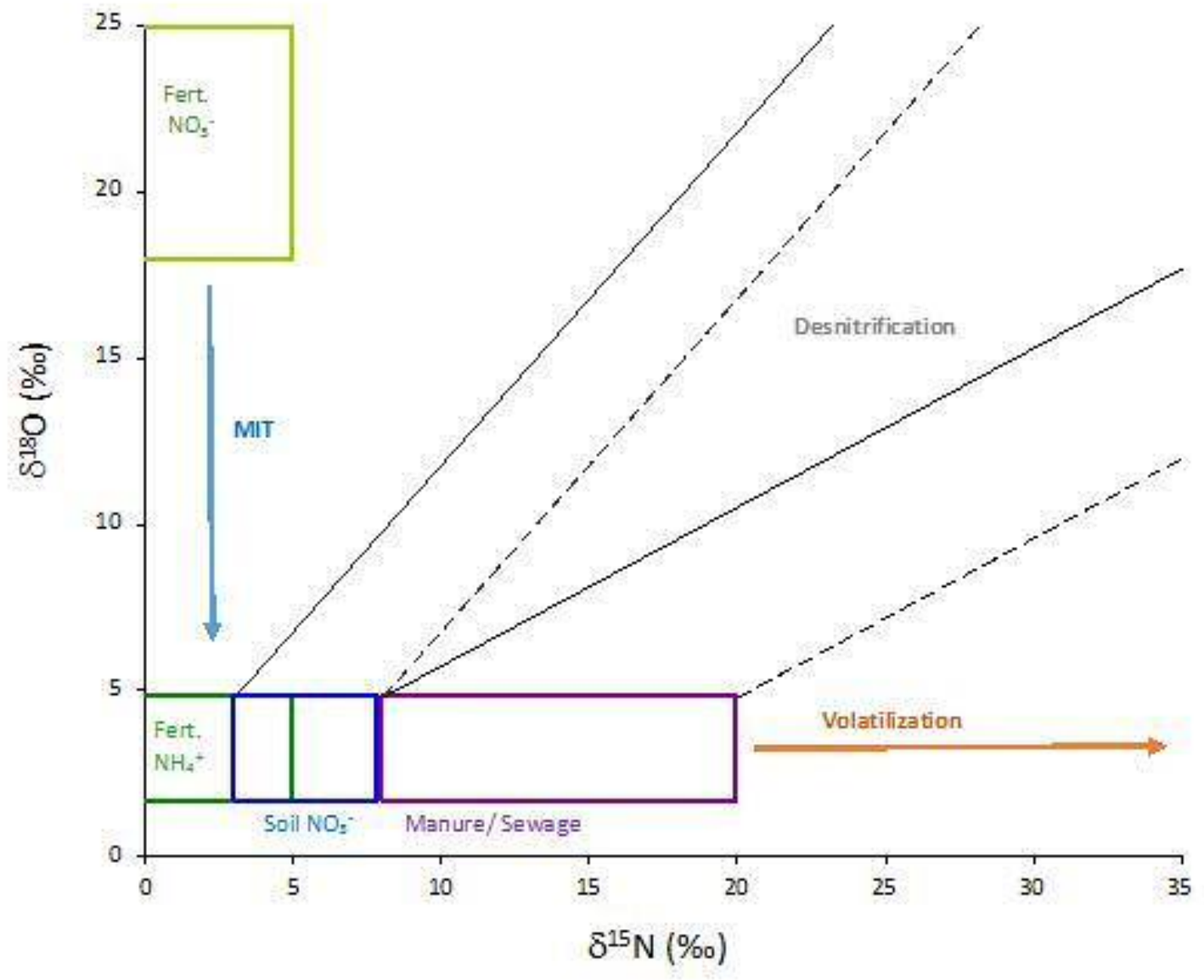

Figure S10. $\delta^{15}N$–$NO_3^-$ **and** $\delta^{18}O$–$NO_3^-$ showing the compositional boxes according to nitrate origin. Fertilizer data are from Vitòria et al. (2004), organic fertilizers and sewage data are from Widory et al. (2004), soil nitrate data are from Mengis et al. (2001). **The** $\delta^{18}O$–$NO_3^-$ of nitrification was calculated using the maximum and minimum $\delta^{18}O$–$H_2O$ values in the studied area (Eq. 1). Arrows indicate theoretical evolution during volatilization and MIT. The theoretical denitrification areas were calculated **using published εN:εO slopes of 2:1**(Böttcher et al., 1990) and 1.3:1(Fukada et al., 2003). The range of $\delta^{18}O$–$NO_3^-$ theoretical values for nitrate from ammonium nitrification has been calculated using bibliographical values of groundwater of Mediterranean influence (Araguas-araguas, 2005), $\delta^{18}O$–$H_2O$ oscillates between **–9.0 ‰** and **–4.3 ‰, giving rise to values of** $\delta^{18}O$–$NO_3^-$ **between + 1.7 ‰ and + 4.8 ‰ (see Eq** 1).

***Boron***

The values of $\delta^{11}B$ in natural waters (seawater, surface, groundwater, geothermal fluids, brines) have a range of values from **–16 ‰ to + 60 ‰** (Tirez et al., 2010). This wide range allows the boron isotopic composition to be used as a tracer of pollutants in groundwater. The $\delta^{11}B$ of many contaminants such as wastewater, landfill leachate, and agricultural irrigation returns show different isotopic compositions that can help to identify the source of contamination (Vengosh et al., 1994; Widory et al., 2005). Some studies have used boron isotopes as a tracer of landfill leachate in groundwater (Nigro et al., 2018). The reported values of urban waste landfill leachates range from -6.0 ‰ to + 25.1 ‰ (Barth, 2000; Hogan and Blum, 2003), different from the isotopic composition of the seawater boron (+ 39.5 ‰).

Water from wastewater treatment plants or industrial effluents greatly affects boron in groundwater and surface water (Chetelat and Gaillardet, 2005; Pennisi et al., 2006; Vengosh et al., 1994). However, Neal et al. (Neal et al., 2010) reported a decrease in boron content in rivers and wastewater treatment plant effluents. due to the substitution of percarbonates as a bleaching agent in soaps and detergents. At the same time, in some areas, boron derived from urban effluents has increased its $\delta^{11}B$ to values around +12 ‰ (Guinoiseau et al., 2018). That is why a good characterization of the boron derived from these effluents is necessary to correctly map out the contribution B of sources (Figure S9).

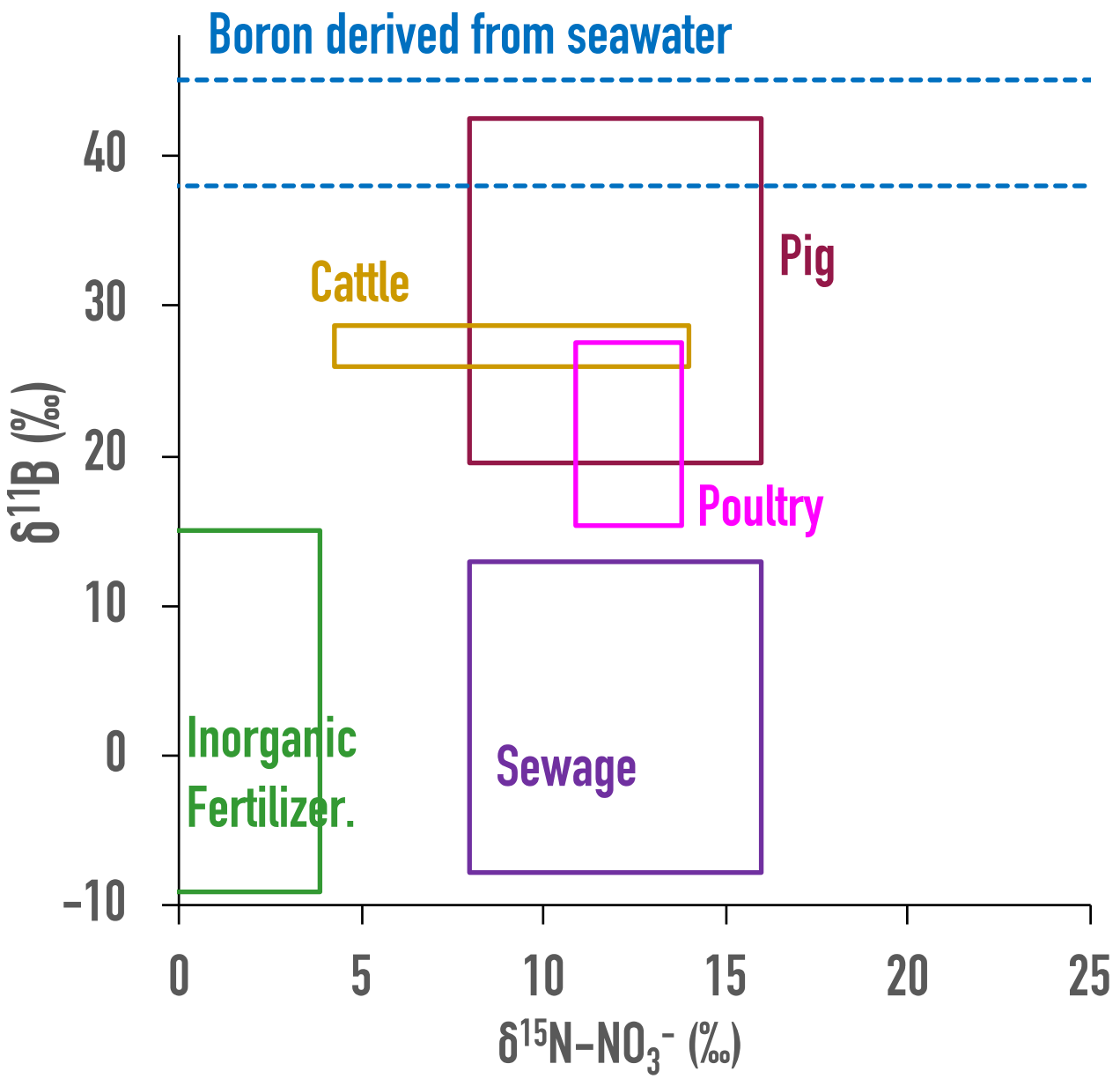

Figure S11. Graph of $\delta^{11}B$ vs. $\delta^{15}N$ with the various compositional boxes derived from bibliographic values (Seiler, 2005; Widory et al., 2004; Widory et al., 2005)